\documentclass[11pt]{article}
\usepackage[margin=1in]{geometry}
\usepackage[utf8]{inputenc}

\usepackage{amsmath,amssymb,amsfonts,amsthm}

\usepackage[font=small,labelfont=bf]{caption}

\usepackage{breqn}

\usepackage{bold-extra}

\usepackage{bm}

\usepackage{graphicx}
\usepackage[dvipsnames,svgnames]{xcolor}
\usepackage{url}
\usepackage{hyperref}
\hypersetup{
  colorlinks   = true, 
  urlcolor     = blue, 
  linkcolor    = blue, 
  citecolor   = blue 
}

\usepackage[numbers,sort&compress]{natbib}
\let\OLDthebibliography\thebibliography
\renewcommand\thebibliography[1]{
  \OLDthebibliography{#1}
  \setlength{\parskip}{4pt}
  \setlength{\itemsep}{0pt plus 0.3ex}
}

\allowdisplaybreaks
\usepackage{tabularx}
\usepackage{bbm}
\usepackage{bm}
\usepackage{listings}
\usepackage{longtable}
\usepackage{titlesec}
\titleformat*{\section}{\bfseries\boldmath}
\titleformat*{\subsection}{\bfseries\boldmath}
\titleformat*{\subsubsection}{\bfseries\boldmath}

\usepackage{array}

\usepackage{multicol}
\usepackage{enumitem}

\usepackage[frozencache,cachedir=minted-cache]{minted}
\definecolor{bg}{rgb}{0.95,0.95,0.95}

\newcommand\refeq[1]{Eq.~(\ref{#1})}

\newcommand\refta[1]{Tab.~\ref{#1}}
\newcommand\refse[1]{Sect.~\ref{#1}}

\newcommand\citere[1]{Ref.~\cite{#1}}
\newcommand\citeres[1]{Refs.~\cite{#1}}
\newcommand\refap[1]{App.~\ref{#1}}
\def\reffi#1{\mbox{Fig.~\ref{#1}}}

\newcommand{\htb}[1]{{\color{black} #1}}

\newcommand{\htmm}[1]{{\color{black} #1}}

\newcommand{\hto}[1]{{\color{black} #1}}
\newcommand{\htg}[1]{{\color{black} #1}}
\newcommand{\htbr}[1]{{\color{black} #1}}

\newcommand{\jmn}[1]{{\color{black} #1}}

\newcommand{\TBa}[1]{{\color{black} #1}}

\usepackage{soul}

\newcommand{\KRn}[1]{\textcolor{black}{#1}}

\newcommand{\GW}[1]{\textcolor{black}{#1}}
\newcommand{\GWn}[1]{\textcolor{black}{#1}}

\newcommand{\TBn}[1]{\textcolor{black}{#1}}

\newcommand{\gev}{\ \mathrm{GeV}}

\graphicspath{{Plots/}}

\begin{document}

\thispagestyle{empty}

\def\thefootnote{\fnsymbol{footnote}}

\begin{flushright}
  \footnotesize
  DESY-23-144 \qquad KA-TP-19-2023 \qquad IFT-CSIC-120
\end{flushright}

\vspace*{0.4cm}

\begin{center}


{\Large \textbf{First shot of the smoking gun: probing the electroweak\\[0.4em] phase transition in the 2HDM with novel searches\\[0.5em] for $A \to ZH$ in $\ell^+ \ell^- t \bar t$ and $\nu \nu b \bar b$ final states}}


\vspace{1.8em}
   


Thomas Biek\"otter$^1$\footnote{thomas.biekoetter@kit.edu},
Sven Heinemeyer$^2$\footnote{sven.heinemeyer@cern.ch},
Jose Miguel No$^{2,3}$\footnote{josemiguel.no@uam.es},\\[0.4em]
Kateryna Radchenko$^4$\footnote{kateryna.radchenko@desy.de},
Mar\'ia Olalla Olea Romacho$^5$\footnote{mariaolalla.olearomacho@phys.ens.fr}
and
Georg Weiglein$^{4,6}$\footnote{georg.weiglein@desy.de}

\vspace*{0.4em}

\textit{
$^1$Institute for Theoretical Physics,
Karlsruhe Institute of Technology,\\
Wolfgang-Gaede-Str.~1, 76131 Karlsruhe, Germany\\[0.2em]
$^2$Instituto de Física Teórica UAM-CSIC, Cantoblanco, 28049,
Madrid, Spain\\[0.4em]
$^3$Departamento de F{í}sica Te{ó}rica, Universidad
Aut{ó}noma de Madrid (UAM),\\
Campus de Cantoblanco, 28049 Madrid, Spain\\[0.2em]
$^4$Deutsches Elektronen-Synchrotron DESY,
Notkestr.~85, 22607 Hamburg, Germany\\[0.2em]
$^5$Laboratoire de Physique de l’Ecole Normale Sup\'erieure, ENS,\\
Universite PSL, CNRS, Sorbonne Universit\'e, Universit\'e Paris Cit\'e,
F-75005 Paris, France\\[0.2em]
$^6$II.~Institut f\"ur Theoretische Physik, Universität Hamburg,\\
Luruper Chaussee 149, 22607 Hamburg, Germany
}

\vspace*{0.2cm}

\begin{abstract}
Recently the ATLAS collaboration has reported the first results of searches for heavy scalar resonances decaying into a $Z$ boson and a lighter new scalar resonance, where the $Z$ boson decays leptonically and the lighter scalar decays into a top-quark pair, giving rise to~$\ell^+ \ell^- t \bar t$ final states. This had previously been identified as a {\it smoking-gun} signature at the LHC for a first-order electroweak phase transition (FOEWPT) within the framework of two Higgs doublet models~(2HDMs). 
In addition, ATLAS also presented new limits where the $Z$ boson decays into pairs of neutrinos and the lighter scalar resonance into bottom-quark pairs, giving rise
to the $\nu \nu b \bar b$ final state. We analyze the impact of these new searches
on the 2HDM parameter space, with emphasis on their capability to probe currently allowed 2HDM regions featuring a strong FOEWPT. We also study the complementarity of these new searches with other LHC probes that could target the FOEWPT region of the 2HDM. Remarkably, the ATLAS search in the $\ell^+ \ell^- t \bar t$ final state shows a local $2.85\,\sigma$ excess 
(for masses of about 650~GeV and 450~GeV for the heavy and light resonance) in the 2HDM parameter region that would yield a FOEWPT in the early universe, which could constitute the first experimental hint of baryogenesis at the 
electroweak scale. We analyze the implications of this excess, and discuss the detectability prospects for the associated gravitational wave signal from the FOEWPT.
Furthermore, we project the sensitivity reach of the 
$\ell^+ \ell^- t \bar t$ signature 
for the upcoming runs of the LHC.
Finally, we introduce the python package \texttt{thdmTools}, a state-of-art tool for the exploration of the 2HDM.
\end{abstract}

\end{center}

\renewcommand{\thefootnote}{\arabic{footnote}}
\setcounter{footnote}{0} 

\newpage

{
  \hypersetup{linkcolor=black}
  \tableofcontents
}

\section{Introduction}
\label{sec:intro}

The Standard Model~(SM) of particle physics predicts the existence of a fundamental
scalar particle as a consequence of the Higgs mechanism. In the year 2012 a scalar particle
with a mass of about~125~GeV was discovered at the Large Hadron Collider~(LHC)~\cite{ATLAS:2012yve,CMS:2012qbp}.
So far, the properties of the detected scalar 
are compatible with the predictions of the~SM.
However, the experimental precision of coupling measurements of the detected Higgs boson is still at the level of~10\% at best~\cite{CMS:2022dwd,ATLAS:2022vkf}, such that there is ample room for interpretations of 
the detected Higgs boson in various theories Beyond the Standard Model~(BSM). 

The~SM predictions have been tested at the LHC at various energy scales, and
so far the measurements are in remarkable agreement with
the SM~\cite{ATLAS:2022djm}. On the other hand, the SM fails to address various major existing observations in Nature.
One of the most pressing open questions in this context concerns
the origin of the matter-antimatter asymmetry of the Universe, which (according to the measured value of the mass of the Higgs boson) cannot be explained in the~SM~\cite{Kajantie:1996mn}.
BSM theories can address the shortcomings of the SM. In particular, models
featuring extended Higgs sectors could allow for the generation of the
baryon asymmetry of the Universe~(BAU) via electroweak (EW) baryogenesis~\cite{Kuzmin:1985mm}.
One of the simplest constructions to realize EW baryogenesis 
is based on a Higgs sector containing a second Higgs 
doublet~\cite{Lee:1973iz,Kim:1979if,Wilczek:1977pj}, i.e.~the Two Higgs Doublet Model~(2HDM).
This model contains two CP-even Higgs bosons, $h$ and $H$, one CP-odd Higgs boson, $A$ and a pair of charged Higgs bosons, $H^\pm$. In the 2HDM the phase transition
giving rise to EW symmetry breaking can be rendered
to be a sufficiently strong first-order transition,
providing the out-of-equilibrium conditions required for 
EW baryogenesis~\cite{Cline:1996mga,Fromme:2006cm,
Cline:2011mm,Dorsch:2016nrg}. Since the Higgs potential has to be significantly
altered w.r.t the one of the SM in order to yield a 
first-order EW phase transition (FOEWPT), 
this generally implies the existence of
new physics at the EW scale that can be searched for at the~LHC.\footnote{See Refs.
\cite{Ashoorioon:2009nf,Curtin:2014jma} for counter examples.}
Moreover, a cosmological first-order phase transition would lead to the generation of
a stochastic primordial gravitational wave~(GW) background that could be detectable with
future space-based gravitational wave observatories, such as the
Laser Interferometer Space Antenna~(LISA)~\cite{Caprini:2019egz,LISACosmologyWorkingGroup:2022jok}.

The possibility of accommodating a strong FOEWPT in the~2HDM has been studied
abundantly in the past (see e.g.~\cite{Cline:1996mga, Fromme:2006cm,Dorsch:2013wja,Dorsch:2014qja, Dorsch:2016nrg, 
Basler:2016obg,Dorsch:2017nza,Goncalves:2021egx,Biekotter:2022kgf}).
It was found that sizable quartic couplings between the scalar states
are required to generate a radiatively and
thermally induced potential barrier between
the symmetry-conserving and the symmetry-breaking
vacua of the 2HDM Higgs potential, thus
facilitating the presence of a FOEWPT.
In addition, assuming that the lightest Higgs
boson~$h$ is the one that is identified with the detected Higgs
boson at~125~GeV, the strength of the transition is maximized (for other parameters fixed)~\cite{Dorsch:2017nza} in
the so-called \textit{alignment limit} of the 2HDM, where the properties of the state~$h$ resemble the ones of the Higgs boson as predicted by the~SM (see \refse{sec:2hdm} for a more detailed discussion), and which is accordingly also favored by the measurements of the signal rates of the Higgs boson at~125~GeV. The sizable quartic couplings required to
facilitate the FOEWPT imply, in combination with other experimental and theoretical restrictions,
that the strongest phase transitions occur in scenarios with relatively large mass splittings between the BSM Higgs bosons~\cite{Dorsch:2014qja,Goncalves:2021egx,Biekotter:2022kgf}, in particular characterized by a mass spectrum $m_H \ll m_A \approx m_{H^\pm}$.
As a consequence of this mass splitting between $H$ and $A$ the decay
$A \to ZH$ is kinematically open~\cite{Dorsch:2014qja}, which (in contrast to the decay $A \to Zh$) remains unsuppressed in the alignment limit of the~2HDM.
Due to this feature, together with the result that a larger
cross section for the process $pp \to A \to ZH$ is correlated with a stronger phase transition~\cite{Biekotter:2021ysx,Biekotter:2022kgf},
this process has been coined a \textit{smoking gun signature} for a FOEWPT at the~LHC~\cite{Dorsch:2014qja,Biekotter:2022kgf}.

Independently of the nature of the EW phase transition, 
the fact that two BSM states are involved in the process $A \to ZH$ 
implies that searches for this decay are of particular phenomenological
importance in the 2HDM. It should be noted that
measurements of the EW precision observables constrain the amount
of weak isospin breaking induced by the second Higgs doublet, enforcing approximate
mass degeneracy between either $H$ and $H^\pm$
or $A$ and $H^\pm$~\cite{Gunion:1989we}.\footnote{Regarding the experimental value 
of the $W$-boson mass that is used in this context, we refer to the combined value excluding the CDF measurement, see \citere{Amoroso:2023pey} for a detailed discussion. Conversely, using the combined value including the CDF measurement quoted in \citere{Amoroso:2023pey} yields a preference for a non-vanishing mass splitting
between $H^\pm$ and the neutral BSM states $H,A$,
see e.g.\ \citeres{Lu:2022bgw,Bahl:2022xzi}. However, we do not consider this possibility here.}
Accordingly, a possible detection of a $A \to ZH$ signal 
would provide information about the entire mass spectrum of the 2HDM. Furthermore, the cross section corresponding to the hypothetical signal would rather precisely
fix the otherwise free parameter $\tan\beta$ (defined as the
ratio of the vacuum expectation values~(vev)
of the Higgs fields in the considered type of the 2HDM), which controls
the couplings of the BSM scalars to fermions and gauge bosons.
Assuming CP conservation in the Higgs sector,
the only remaining free parameter of the~2HDM is then the soft mass parameter
$m_{12}^2$, which would however be restricted
to a specific interval based on theoretical constraints from vacuum
stability and perturbativity~\cite{Arco:2022jrt}. 
\KRn{\TBn{Further constraints} on the
\TBn{2HDM} 
parameter space
\TBn{can be obtained}
from flavor\TBn{-physics} observables.
\TBn{In particular, the measurements of
transition rates of radiative $B$-meson
decays give rise to a lower limit on
the mass of the charged scalars
in the type~II and the type~IV.
Using the current experimental world average
value $\mathrm{BR}(B \to X_s \gamma) =
3.49 \pm 0.19$ from the HFLAV
collaboration~\cite{HFLAV:2022esi}, one finds a limit of
$m_{H^\pm} \gtrsim 500\gev$~\cite{Misiak:2020vlo, Steinhauser}.\footnote{We
note that the
2022 HFLAV average value for $\mathrm{BR}(B \to X_s \gamma)$
used here does not include the most recent Belle-II
result~\cite{Belle-II:2022hys},
which is anyway in good agreement with the
current HFLAV average value.}}
The application of \TBn{the 
flavor-physics constraints} relies on the
assumption that there are no other BSM
contributions to the flavor observables in addition to those of the new 2HDM scalars. 
As such, \TBn{these constraints} 
should be regarded as complementary to the constraints
from direct searches at colliders,
where the latter can be considered as more direct
exclusions that are largely independent of
other possibly existing BSM effects.
Therefore, in this paper, we focus on the
constraints from collider searches and
do not carry out a 
detailed evaluation of the exclusion
limits from flavor-physics observables
\TBn{and their complementarity with
the limits from direct searches}.}

Searches for the decay $A \to ZH$ have been performed by both the ATLAS and the CMS
collaboration utilizing the 8~TeV and 13~TeV data sets. 
These searches made use of the leptonic decay modes of
the $Z$~boson, while for the lighter BSM resonance~$H$
decays either into bottom-quark, tau-lepton or $W$-boson 
pairs were considered~\cite{Khachatryan:2016are,ATLAS:2018oht,Sirunyan:2019wrn,ATLAS:2020gxx}.
Notably, until recently no searches for the $A \to ZH$ decay existed assuming the
decay of~$H$ into top-quark pairs, which are favored for low $\tan\beta$. 
Based on the combination 
of LHC searches and other experimental constraints, however, in the type~II
and type~IV 2HDM (see Sect.~\ref{sec:2hdm}) the allowed region for the
scalar spectrum is such that typically $H$ is heavier than twice the top-quark
mass, which implies that the decay $H \to t \bar t$ is
dominant (except for parameter regions with a large enhancement of the couplings to bottom quarks). 
It was therefore pointed out~\cite{Biekotter:2021ysx,Goncalves:2022wbp,Biekotter:2022kgf} that the searches for $A \to ZH \to \ell^+ \ell^- t \bar t$ hold great promise to probe so far unconstrained parameter space regions of the 2HDM, in particular, as discussed above, the regions suitable for a realization of a FOEWPT.\footnote{See also \citeres{Goncalves:2021egx,MammenAbraham:2022yxp} for recent discussions of the smoking gun signature and its possible impact on the 2HDM based on expected sensitivities at the LHC. }

Recently, the experimental situation has changed with the first public results of searches for 
the signature $A \to ZH \to \ell^+ \ell^- t \bar t$ by the ATLAS collaboration utilizing the
full Run~2 dataset collected at~13~TeV~\cite{ATLAS-CONF-2023-034}.
In addition, ATLAS also presented new searches using
the decay of the $Z$ boson into pairs of neutrinos and 
assuming the decay of the lighter scalar resonance into bottom-quark pairs, giving rise
to the $\nu \nu b \bar b$ final state. \KRn{CMS has not yet released a result in the $\ell^+ \ell^- t \bar t$ final state but the experimental analysis is ongoing~\cite{dpgcms,fischerthesis}.}
In this work we will demonstrate that already the 
first released LHC experimental result on searches in the 
$A \to ZH \to \ell^+ \ell^- t \bar t$ channel
has a large impact on the parameter space of extended Higgs sectors 
and on possible scenarios giving rise to a strong FOEWPT, 
constraining sizable regions of so far allowed 2HDM parameter space.
Besides, we discuss the possible phenomenological and cosmological implications of an excess of events with $2.85\,\sigma$ (local) significance at masses of 450~GeV and 650~GeV for the lighter and heavier BSM resonances, respectively,
reported by ATLAS in this search and compatible with a strong FOEWPT in the 2HDM. 
Furthermore we analyze the strength of the GW signals in the parameter space that is compatible with the observed excess, and discuss the future detectability by LISA.
In addition, we
identify another potentially promising LHC search to target the region indicative
for strong FOEWPT, namely the production of the charged Higgs $H^\pm$, then decaying as 
$H^\pm \to W^\pm H \to \ell^\pm \nu t \bar t$, for which so far no results exist.

Performing an extrapolation of the 95\% confidence-level expected cross section
limits based on the LHC Run~2 presented by ATLAS in~\cite{ATLAS-CONF-2023-034}, 
we also investigate the future discovery reach of the smoking gun search.
If this search were to lead to the detection of a signal, it will be rather straightforward to assess whether this signal can be interpreted within the context of the 2HDM.
If so one can infer the allowed ranges
of the 2HDM parameters according to the discussion above and make predictions
about the associated phenomenology regarding
other LHC searches and the nature of the EW phase transition.
On the other hand, if a signal is observed that cannot be accommodated by 
the 2HDM (e.g.~because the required mass splitting is too large), 
this would be a clear indication for physics beyond the~2HDM. Economical extensions
would be, for instance, the complex (CP-violating) 2HDM or singlet-extended 2HDMs,
where the cross sections for the smoking gun signature and the mass splittings between
the scalars are less restricted as a consequence
of additional parameters~\cite{Biekotter:2021ysx,Biekotter:2021ovi}.
In this manner, the detection of a signal in the $A \to ZH$ 
channel would provide invaluable information about the discrimination between
different scenarios featuring extended scalar sectors and, therefore,
about the underlying physics governing the dynamics of EW symmetry breaking.

The outline of this paper is as follows: In \refse{sec:theo} we introduce the 2HDM and
specify our notation. Moreover, we briefly summarize the basis for the description
of the EW phase transition in the early universe with regards to strong FOEWPTs,
EW baryogenesis and GWs. 
In \refse{sec:numanal}
we present the numerical discussion of the impact of the new ATLAS $A \to ZH$ searches
on the 2HDM parameter space. We divide our discussion of the new constraints
into two parts, focusing on a low-$\tan\beta$ and a high-$\tan\beta$ regime
in \refse{sec:lowtanbeta} and \refse{sec:hightanbeta}, respectively.
Then, \refse{sec:prospects} is devoted to an evaluation of the future prospects of
the $\ell^+ \ell^- t \bar t$ searches. The excess observed by ATLAS in the $\ell^+ \ell^- t \bar t$ final state as a hint of a FOEWPT in the 2HDM is analysed in \refse{sec:excess}, and its connection to the possible existence of a stochastic GW signal from the EW phase transition is discussed.
We summarize our findings and conclude in \refse{sec:conclus}.
\refap{app:code} is devoted to presenting the python package \texttt{thdmTools}, a state-of-art tool for the exploration of the 2HDM which we have used for our analyses.
\refap{app:proj} contains a comparison of the
future projections obtained here based on the
new ATLAS results to earlier projections based on
expected cross-section limits from~CMS.

\section{Theoretical background}
\label{sec:theo}
In order to specify our notation, we provide an overview of the 2HDM Higgs sector below.
We will also briefly discuss the methods used to investigate the occurrence of a strong FOEWPT, and the related phenomenological consequences, such as the realization of EW baryogenesis or the generation of a primordial GW background.

\subsection{The Two Higgs Doublet Model (2HDM)}
\label{sec:2hdm}
As theoretical framework for our study we consider an extension of the SM Higgs sector by one additional complex SU(2) doublet.~Models 
with two Higgs doublets have been widely studied in the literature 
\cite{Lee:1973iz,Branco:2011iw} and give rise to a rich phenomenology including the possibility of a 
strong FOEWPT. In the present analysis we assume a CP-conserving 2HDM with a softly broken $\mathbb{Z}_2$ 
symmetry. This discrete symmetry between the doublets leaves the potential invariant
(up to the soft breaking) under transformations
of the type $\Phi_1 \rightarrow \Phi_1 ,\; \Phi_2 \rightarrow -\Phi_2$ in order to avoid the occurrence of large flavour changing neutral currents that 
would violate the existing bounds. We allow for a soft $\mathbb{Z}_2$ -breaking via a mass parameter $m_{12}^2$. Given these constraints,
the most general form of the Higgs potential is
\begin{eqnarray}
V(\Phi_1,\Phi_2) &=& m_{11}^2 (\Phi_1^\dagger\Phi_1) + m_{22}^2 (\Phi_2^\dagger\Phi_2)
- m_{12}^2 (\Phi_1^\dagger \Phi_2 + \Phi_2^\dagger\Phi_1)
+ \frac{\lambda_1}{2} (\Phi_1^\dagger \Phi_1)^2 +
\frac{\lambda_2}{2} (\Phi_2^\dagger \Phi_2)^2 \nonumber \\
&& + \lambda_3
(\Phi_1^\dagger \Phi_1) (\Phi_2^\dagger \Phi_2) + \lambda_4
(\Phi_1^\dagger \Phi_2) (\Phi_2^\dagger \Phi_1) + \frac{\lambda_5}{2}
[(\Phi_1^\dagger \Phi_2)^2 +(\Phi_2^\dagger \Phi_1)^2],
\label{eq:scalarpot}
\end{eqnarray}
where the two doublets are conveniently parameterized as
\begin{eqnarray}
\Phi_1 = \left( \begin{array}{c} \phi_1^+ \\ \frac{1}{\sqrt{2}} (v_1 +
    \rho_1 + i \eta_1) \end{array} \right) \;, \quad
\Phi_2 = \left( \begin{array}{c} \phi_2^+ \\ \frac{1}{\sqrt{2}} (v_2 +
    \rho_2 + i \eta_2) \end{array} \right) \; .
\label{eq:2hdmvevs}
\end{eqnarray}
Here $v_i$ is the vacuum expectation value (vev) acquired by the respective doublet. The eight degrees of freedom $\phi_i^{+}, \rho_i$ and $\eta_i$ mix to give rise to the massive scalars $h$, $H$ (CP-even), $A$ (CP-odd),
$H^{\pm}$ and the three Goldstone bosons $G^0, G^{\pm}$. In this work we identify $h$ with the Higgs boson at 125 GeV detected at the LHC. The rotation from the gauge basis to the mass basis involves two mixing matrices with the mixing angles $\alpha$ and $\beta$ 
for the CP-even and the CP-odd/charged sector, respectively.

The coefficients $m_{11}^2,\; m_{22}^2,\; m_{12}^2,\; \lambda_i\; (i=1,...5)$ in~\refeq{eq:scalarpot} can be mapped to the physical basis of the masses and mixing angles. We employ the usual convention of the parametrization of the mixing angles as $\tan\beta = v_2/v_1 $, and
$\cos(\beta-\alpha)$.
In the limit $\cos (\beta-\alpha) \rightarrow 0$ -- the \textsl{alignment} limit --~\cite{Gunion:2002zf}
the light Higgs boson $h$ in the 2HDM has the same tree-level couplings to the SM fermions and gauge bosons as the SM Higgs boson.
In the numerical discussion, we use the following basis of free parameters of the model,
\begin{equation}
\cos (\beta-\alpha) \; , \quad \tan\beta \;, \quad v \; ,
\quad m_h\;, \quad m_H \;, \quad m_A \;, \quad m_{H^{\pm}} \;, \quad m_{12}^2\;,
\label{eq:freeparas}
\end{equation}
where two of these parameters are already fixed by experiment, namely $m_h \approx 125 \gev$ and
$v=\sqrt{v_1^2+v_2^2} \approx 246\gev$.

Extending the 
$\mathbb{Z}_2$ symmetry to the Yukawa sector
leads to four different 2HDM types depending on the $\mathbb{Z}_2$ parities
of the fermions, summarized in \refta{tab:yuktypes}. It should be noted that due to the opposite $\mathbb{Z}_2$ parities
of the Higgs doublets each fermion can only be coupled to either $\Phi_1$ or $\Phi_2$.
In \refta{tab:yuktypes} we also give the $\tan\beta$ dependence of the coupling modifyers
of the CP-odd Higgs boson $A$ of the 2HDM 
to up-type and down-type quarks, $\xi_A^{u,d}$ (see \citere{Arco:2022xum} for a formal definition of these parameters).
To a very good approximation the cross sections for the production of~$A$ at the LHC are
proportional to the square of the coupling modifyers $\xi_A^{u,d}$. 
Since $\xi_A^u = \cot\beta$ for all types, the dominant production mode
for values of $\tan\beta \approx 1$ is gluon-fusion production involving
a top-quark loop. For values of $\tan\beta \gtrsim 10$ the production in association with
bottom-quark pairs becomes important in type~II and type~IV, in which $\xi_A^{d} = \tan\beta$, whereas in type~I and type~III
this production mode is always substantially smaller than gluon-fusion production. 

\begin{table}[t]
\centering
    \begin{tabular}{l|lll|rr}
    	& u-type     & d-type     & leptons   & $\xi_A^u$    & $\xi_A^d$\\
    	\hline
    	type I                     & $\Phi_{2}$ & $\Phi_{2}$ & $\Phi_{2}$ & $\cot\beta$  &  $-\cot\beta$ \\
    	type II                    & $\Phi_{2}$ & $\Phi_{1}$ & $\Phi_{1}$ & $\cot\beta$  &  $\tan\beta$ \\
    	type III (lepton-specific) & $\Phi_{2}$ & $\Phi_{2}$ & $\Phi_{1}$ & $\cot\beta$  &  $-\cot\beta$ \\
    	type IV (flipped)          & $\Phi_{2}$ & $\Phi_{1}$ & $\Phi_{2}$ & $\cot\beta$  &  $\tan\beta$
    \end{tabular}
    \caption{Couplings of the two Higgs doublets
    to the SM fermions in the four types of
    the~2HDM. Right: $\tan\beta$-dependence of
    the couplings of the up-type and
    down-type quarks to the CP-odd Higgs boson $A$.}
       \label{tab:yuktypes}
\end{table}

\subsection{Thermal history analysis}
\label{sec:thermal}

In order to study the physics of the EW phase transition, we will use the finite-temperature effective potential formalism. The one-loop, daisy resummed, finite-temperature 2HDM effective potential is 
given as
\begin{equation}
    V_{\rm eff} = V_{\rm tree} + V_{\rm CW} +
        V_{\rm CT} + V_{\rm T} + V_{\rm daisy} \ .
    \label{eq:fullpotential}
\end{equation}
The temperature-independent part of the potential comprises the first three terms, where $V_{\rm tree}$ is defined
in \refeq{eq:scalarpot}, $V_{\rm CW}$ is the one-loop Coleman Weinberg potential~\cite{Coleman:1973jx} incorporating
the radiative corrections, and $V_{\rm CT}$ is a UV-finite counterterm potential introduced in order to keep the physical masses and the vevs of the Higgs fields at their tree-level values at zero temperature~\cite{Basler:2016obg}.
The thermal corrections to the scalar potential are split into two terms.
The first one, $V_{\rm T}$,
incorporates the one-loop thermal corrections in terms of the well-known $J$-functions (see e.g.~\citere{Quiros:1999jp}).
The second term, $V_{\rm daisy}$, is an additional piece accounting for the
resummation of the so-called daisy-diagrams which signal the breakdown of fixed-order perturbation theory at finite temperature.
As resummation prescription, we follow the Arnold-Espinosa
method~\cite{Arnold:1992rz}, which resums the
infrared-divergent contributions from
the bosonic Matsubara zero-modes. \jmn{We emphasize that 
the  computation of the finite-temperature effective potential, at the order performed in this work, 
is affected by sizable theoretical uncertainties, see~\cite{Curtin:2016urg,Croon:2020cgk,Gould:2021oba,Gould:2023ovu} for a detailed discussion.}
\GW{As a consequence,} the regions studied in the following should only be regarded as indicative for the presence of a \GW{strong} FOEWPT. 

\subsection{Vacuum tunneling}

In order to find regions in the parameter space of the 2HDM
that feature a FOEWPT we track the co-existing vacua as a function of the temperature 
using the effective potential from~\refeq{eq:fullpotential},
by means of a modified version of the public code \texttt{CosmoTransitions}~\cite{Wainwright:2011kj}.
Typically, the universe evolves starting from
an EW symmetric vacuum configuration at the origin of field
space.\footnote{EW symmetry 
non-restoration in the high-temperature regime $T \gg m_A, m_{H^{\pm}}, m_H, M$ (with $ M^2 \equiv m_{12}^2/(s_{\beta}\,c_{\beta})$)
is possible in the 2HDM~\cite{Biekotter:2022kgf}. For the
sake of simplicity, we omit this possibility 
in the discussion of FOEWPT here.}
We identify the 2HDM parameter space regions which, as the universe cools down, feature an EW-breaking minimum of the Higgs potential that
is separated from the minimum in the origin by a potential barrier.
The universe reaches 
the critical temperature $T_c$ when these two coexisting vacua are degenerate. At later times, when $T < T_c$, the minimum corresponding to the EW vacuum drops below the minimum in the origin, and thus becomes energetically more favorable.
At this point, the onset of the first-order phase transition from the 
minimum at the origin to the EW
vacuum depends on the transition rate per unit time and unit volume.
The transition rate in turn depends on the temperature-dependent Euclidean bounce action $S(T)$ of the (multi-)scalar 
field configurations. The onset of the phase transition occurs when (see e.g.~\citere{LISACosmologyWorkingGroup:2022jok})
\begin{equation}
    S(T_n)/T_n \approx 140 \ ,
    \label{eq:onset}
\end{equation}
which arises from the comparison of the transition rate and the expansion rate of the universe.
$T_n$ is the nucleation temperature, which very accurately corresponds to the temperature at which the transition takes place.
If the condition~\eqref{eq:onset} is not fulfilled at any temperature $T < T_c$,
the phase transition cannot complete successfully, and the universe remains trapped in the false
vacuum at the origin~\cite{Biekotter:2022kgf} (see also~\citeres{Baum:2020vfl, Biekotter:2021ysx}).

\subsection{\htmm{Strong} first-order electroweak phase transitions and baryogenesis}

The origin of the baryon asymmetry of the universe (BAU) is one of the main open questions of modern 
particle
physics.
In extensions of the SM such as the 2HDM it is possible to dynamically generate an excess of
matter over antimatter in the universe
by means of EW baryogenesis.
According
to the Sakharov conditions~\cite{Sakharov:1967dj},
a first-order EW phase transition is a necessary ingredient for the realization of EW baryogenesis
as it provides the required conditions to bring the thermal plasma out of equilibrium.
In order to avoid the washout of the asymmetry after the phase transition,
the EW vev after the transition should be larger than the transition temperature, i.e. 
\begin{equation}
\label{eq:SFOEWPT}
    \xi_n=\frac{v_n}{T_n} \gtrsim 1 \ ,
\end{equation}
%
%
where $v_n$ is the EW vev at the nucleation temperature $T_n$. 
The above condition~\eqref{eq:SFOEWPT}, which defines a strong FOEWPT, yields the parameter region satisfying baryon-number preservation~\cite{Dimopoulos:1978kv} after the transition, 
which is therefore
suitable for EW baryogenesis.
It should be noted that a successful realization of EW baryogenesis also requires BSM sources of
CP violation.
In the present paper we restrict ourselves to the condition for a strong FOEWPT 
and do not carry out a detailed investigation of the actual realization of EW 
baryogenesis \htmm{(see \citere{Goncalves:2023svb}
for a recent study)}. Accordingly,
we make the assumption that the
additional sources of CP violation 
that are needed for EW baryogenesis
do not have a significant impact on the properties of the FOEWPT (see e.g.~\citere{Dorsch:2016nrg} for an example in the 2HDM).

\subsection{Gravitational waves}
\label{sec:gw}

It is well-known that a cosmological first-order phase transition 
gives rise
to a stochastic 
gravitational wave signal~\cite{Witten:1984rs,Hogan:1986qda}. Since the EW phase transition would have happened 
at temperatures comparable to the EW scale, the GW signal spectrum would be largest around milli-Hz frequencies, thus in the best-sensitivity range of the planned LISA space-based GW interferometer~\cite{LISA:2017pwj,LISACosmologyWorkingGroup:2022jok}.
The GWs in a FOEWPT are sourced by the collision of bubbles and the surrounding plasma motions in the form of sound waves~\cite{Hindmarsh:2013xza,Hindmarsh:2015qta,Hindmarsh:2016lnk,Hindmarsh:2017gnf}, as well as the turbulence generated after the collisions~\cite{Caprini:2006jb,Gogoberidze:2007an,Caprini:2009yp,RoperPol:2019wvy,Auclair:2022jod} (see~\citere{Caprini:2019egz} for a review). In the case of the 2HDM, the GW contribution from bubble collisions themselves can be neglected, and the GW power spectrum may be modeled with the sound waves as dominant source~\cite{Dorsch:2016nrg}.
%
There are four phase transition parameters that characterize the corresponding GW signal~\cite{Caprini:2019egz}:
\textit{(i)} the temperature $T_*$ at which the phase transition occurs, which we identify here with the
nucleation temperature $T_n$.\footnote{We could instead consider $T_*$ to be the percolation temperature~\cite{Ellis:2018mja}, at which the phase transition completes from the percolation of bubbles, yet the numerical difference 
compared to $T_n$ is very small.}
\textit{(ii)} the phase transition strength $\alpha$, defined as the difference of
the trace of the energy-momentum tensor
between 
the two vacua involved in the transition,
normalized to the radiation background energy density.
\textit{(iii)} the inverse duration of the
phase transition in Hubble units, $\beta / H$.
\textit{(iv)} the bubble wall velocity in the
rest frame of the fluid (and far from the bubble), $v_{\rm w}$.
To compute $\alpha$ we
follow \citeres{Caprini:2019egz,
LISACosmologyWorkingGroup:2022jok},
\begin{equation}
    \alpha = \frac{1}{\rho_R} \left( \Delta V(T_*) - \left( \frac{T}{4} \frac{\partial \Delta V(T)}{\partial T} \right) \bigg|_{T_*} \right),
    \label{eq:gw_alpha}
\end{equation}
where $\Delta V (T_*)$ is the potential difference between the 
two vacua 
evaluated at the temperature of the phase transition, and
$\rho_R$ is the radiation energy density of the universe. 
The inverse duration of the phase transition $\beta/H$ can be generally calculated as
\begin{equation}
    \frac{\beta}{H} = T_* \left( \frac{d}{dT} \frac{S(T)}{T} \right) \bigg|_{T_*},
    \label{eq:gw_beta}
\end{equation}
%
where $S(T)$ is (as in Eq.~\eqref{eq:onset})
the temperature-dependent (3-dimensional) Euclidean bounce action.
Finally, \jmn{based on recent results indicating that phase transition bubbles preferentially expand with either $v_{\rm w} \approx c_s$ ($c_s$ being the speed of sound of the plasma)\footnote{\jmn{For a relativistic perfect fluid, $c_s = 1/\sqrt{3} \simeq 0.577$.}} or $v_{\rm w} \to 1$~\cite{Laurent:2020gpg,Ai:2023see}} (see also~\citere{Dorsch:2018pat} for a \jmn{further} discussion of bubble wall velocity estimates in BSM theories) we choose to fix $v_{\rm w} = 0.6$ as a representative case. 

Based on the four quantities introduced above, the primordial stochastic GW background produced during a cosmological phase transition can be computed using numerical power-law fits to results of GW production obtained in hydrodynamical simulations of the thermal plasma. In our numerical analysis, we include the contributions to the 
GW power spectrum from sound waves $h^2 \Omega_{\rm sw}$ and turbulence $h^2 \Omega_{\rm turb}$, where sound waves are the dominant GW source for the FOEWPTs considered here.
The specific formulas used in our analysis for the computation of the GW spectral
shapes, their amplitudes and the peak frequencies can be found in \citere{Biekotter:2022kgf}, which closely follows~\citeres{Caprini:2019egz,Auclair:2022jod}.
%
Whether a stochastic GW signal is detectable at a GW 
observatory
depends on the signal-to-noise
ratio~(SNR), 
which can be computed for a specific
parameter point and a specific GW experiment as
\begin{equation}
    {\rm SNR} = \sqrt{\mathcal{T} \int^{+ \infty}_{- \infty} {\rm d}f \left[\frac{h^2\Omega_{\rm GW}(f)}{h^2\Omega_{\rm Sens}(f)}\right]^2},
\end{equation}
where $\mathcal{T}$ is the duration of the experiment, $h^2 \Omega_{\rm Sens}$ is the nominal sensitivity of
the detector, computed according to the mission requirements~\cite{LISAmissionreq},
and $h^2 \Omega_{\rm GW} = h^2 \Omega_{\rm sw} + h^2 \Omega_{\rm turb}$ is the spectral shape of the GW signal.
For the present analysis, we focus on the GW detectability with LISA, for which
we will assume an operation time of $\mathcal{T}$ = 7 years, and consider a GW signal to be detectable if 
SNR~$>1$ (more stringent SNR detection criteria could also be considered~\cite{Caprini:2019egz}).

\section{Numerical analysis}
\label{sec:numanal}

In this section we analyze \jmn{in detail} the impact of the recent ATLAS searches for the $A \to Z H$ 
signature~\cite{ATLAS:2020gxx,ATLAS-CONF-2023-034}
on the 2HDM parameter space. Before we discuss the results of our analysis,
we briefly describe the implementation of the new ATLAS limits into 
\jmn{the~\texttt{HiggsTools} package~\cite{Bahl:2022igd} (which contains \texttt{HiggsBounds}~\cite{Bechtle:2008jh,Bechtle:2011sb,Bechtle:2013wla,Bechtle:2020pkv}),}
discuss the public numerical tools that were used in our analysis and introduce the software package \texttt{thdmTools}
that we have developed as part of this work.

\subsection{Implementation of new limits into \texttt{HiggsBounds}}
\label{sec:hbimple}

In order to \GW{confront} 
the 2HDM \GW{predictions for the different regions of the} 
parameter \jmn{space} 
\GW{with} the new cross section limits reported by the ATLAS Collaboration, we have implemented the \jmn{95\% confidence level (C.L.)} expected
and observed cross section limits into the \texttt{HiggsBounds} dataset, \jmn{corresponding to} 
the following new ATLAS results \GW{that were} not yet contained in the 
\GW{public}
\texttt{HiggsBounds} dataset~\cite{Bahl:2022igd}:
\begin{itemize}
\item[-] $gg \to A \to ZH \to \ell^+ \ell^- b \bar b$ at 13~TeV including $139~\mathrm{fb}^{-1}$
from \citere{ATLAS:2020gxx}
\item[-]
$b \bar b \to A \to ZH \to \ell^+ \ell^- b \bar b$ at 13~TeV including $139~\mathrm{fb}^{-1}$
from \citere{ATLAS:2020gxx}
\item[-]
$gg \to A \to ZH \to \ell^+ \ell^- t \bar t$ at 13~TeV including $\htmm{140}~\mathrm{fb}^{-1}$
from \citere{ATLAS-CONF-2023-034}
\item[-]
$gg \to A \to ZH \to \nu \nu b \bar b$
at 13~TeV including $\htmm{140}~\mathrm{fb}^{-1}$
from \citere{ATLAS-CONF-2023-034}
\item[-]
$b \bar b \to A \to ZH \to \nu \nu b \bar b$
at 13~TeV including $\htmm{140}~\mathrm{fb}^{-1}$
from \citere{ATLAS-CONF-2023-034}
\end{itemize}
We note that the results from \citere{ATLAS:2020gxx} update the previous ATLAS results based on 
\GWn{$36.1~\mathrm{fb}^{-1}$ collected during the first two years}
of Run~2~\cite{ATLAS:2018oht}, which are contained in the 
\GW{public}
\texttt{HiggsBounds} dataset (and which will now be replaced by the full Run~2 results).
The corresponding CMS results include first-year Run~2 data~\cite{Khachatryan:2016are}
and are also implemented in \texttt{HiggsBounds}, but since they are based on less data the extracted limits are weaker than the limits from the new ATLAS \GW{analyses}.

\vspace{1mm}

\GW{In our analysis below we use \texttt{HiggsBounds} to determine the 
parameter regions in the considered 2HDM scenarios that are 
excluded \jmn{at the 95\% C.L.} by the existing limits from Higgs searches. In order to ensure the correct statistical interpretation of the 
excluded regions as a limit at the 95\% C.L., \texttt{HiggsBounds} applies for 
each BSM scalar only the specific observed limit that has the highest 
sensitivity according to the expected limits of all considered searches.
For illustration, we furthermore display the regions that would be excluded by 
different searches if each of these searches were applied individually. In this 
way the impact of the new $A \to ZH$ searches from ATLAS in the different final 
states becomes clearly visible, and one can assess to what extent these 
searches probe parameter regions that were so far unexplored. We note, however, 
that the requirement for a certain parameter point to be simultaneously in 
agreement with the 95\% C.L.\ exclusion limits of all available searches would 
in general result in an excluded region that would be too aggressive in 
comparison to the parameter region corresponding to an overall exclusion at the 
95\% C.L.\ (which we determine using \texttt{HiggsBounds} as described above).}

For the application of the \texttt{HiggsBounds} cross section limits, one has to provide the relevant \GW{model predictions for the} cross sections and branching ratios of the scalar states. To this end, we utilized cross-section predictions from the \GW{corresponding part of the} \texttt{HiggsTools} 
package~\cite{Bahl:2022igd} based on the effective-coupling input (see \citeres{Bahl:2022igd,
LHCHiggsCrossSectionWorkingGroup:2016ypw} and references therein for details), and branching ratios of the Higgs bosons as obtained using 
\texttt{AnyHdecay}~\cite{Djouadi:1997yw,Djouadi:2018xqq,Muhlleitner:2016mzt}.

\jmn{In order to automate the interface to these numerical tools in 2HDM analyses, we have developed} the \GW{public} \texttt{thdmTools} package.
Beyond the interfaces for \texttt{HiggsTools} \hto{(containing \texttt{HiggsBounds} 
and \texttt{HiggsSignals}~\cite{Bechtle:2013xfa,Bechtle:2020uwn,Bahl:2022igd}, where the latter tests the properties of the detected Higgs boson 
at about $125$~GeV against the LHC Higgs boson rate measurements)}
and \texttt{AnyHdecay}, \texttt{thdmTools} also includes additional interfaces for assessing 2HDM parameter points against 
various experimental constraints, including those from electroweak precision observables and flavor physics 
\hto{(where the latter are not explicitly applied \GW{in the following analyses}, see the discussion below)}.
Moreover, \texttt{thdmTools} can be used to check against theoretical 
constraints from 
boundedness from below
and perturbative unitarity. A brief description of the functionalities of \texttt{thdmTools} \jmn{as well as instructions for download and 
installation can be found in} Appendix~\ref{app:tools}.

\subsection{New constraints on the 2HDM parameter space}
\label{sec:newconstr}

\GW{In our analysis we} \jmn{target two separate 2HDM parameter regions:} a low-$\tan\beta$ regime, \jmn{corresponding} to values of
$\tan\beta \leq 3$, and a high-$\tan\beta$ regime with values of $\tan\beta \geq 10$.
In the low $\tan\beta$-regime,~$A$ can be produced with sizable cross sections via gluon-fusion, and we will show that searches in the $\ell^+ \ell^- t \bar t$ final 
state exclude significant parts \jmn{of previously unconstrained parameter space,} whereas the ATLAS searches utilizing $\nu \nu b \bar b$ final states only provide exclusions on the 2HDM parameter space that were already excluded by other LHC searches or theoretical requirements from perturbative
unitarity. In the high-$\tan\beta$ regime, the state~$A$ can be 
produced \GW{with sizable rates} in
$b \bar b$-associated processes \jmn{for the 2HDM of Yukawa type~II and IV.} 
\jmn{In this regime, the $\nu \nu b \bar b$ searches cover regions of parameter space \GW{that have already been} probed via the $\ell^+ \ell^- b \bar b$ final state.}



For intermediate values of $\tan\beta$, in between the low- and high-$\tan\beta$ regimes considered
here, the new ATLAS searches cannot probe substantial parts of the 2HDM parameter space. The reason is that the
gluon-fusion production cross sections of~$A$, dominantly generated via a top-loop diagram, are roughly proportional to~$1 / \tan^2\beta$ for \jmn{intermediate~$\tan\beta$ values,} such that the searches in the $\ell^+ \ell^- t \bar t$ final states quickly loose sensitivity \jmn{as~$\tan\beta$ increases.
At the same time} 
the $b \bar b$-associated production of~$A$ is enhanced in type~II and type~IV by factors of~$\tan^2\beta$, such that the searches relying on this production mechanism become more sensitive for larger values of~$\tan\beta$. \jmn{Yet,} in our numerical analysis we find that one needs roughly an enhancement \jmn{corresponding to $\tan^2\beta \approx 100$ to achieve} cross sections in the $b \bar b$-associated production mode that are comparable to the exclusion limits resulting from the 
\KRn{$gg \to A \to ZH$}
searches. Consequently, we chose the value
$\tan\beta = 10$ as a representative benchmark scenario in order to assess the sensitivity of the new searches
in the high-$\tan\beta$ regime. The impact of the searches for even larger values of~$\tan\beta$ can be extrapolated from
the discussion of this scenario, as will be shown in detail in \refse{sec:hightanbeta}.

\vspace{1mm}

\jmn{As already discussed in~\refse{sec:intro}, the 2HDM parameter region that we explore in this work is motivated by} the possibility of realizing a strong FOEWPT \GW{giving rise to EW baryogenesis}
\jmn{in the early universe.}
%
In this scenario the CP-odd scalar~$A$ and the charged scalars~$H^\pm$ are assumed to be mass-degenerate, i.e.~$m_A = m_{H^\pm}$, and the squared mass scale
$M^2 = m_{12}^2 / (s_\beta c_\beta)$ is set equal to the mass of the
heavy CP-even scalar~$H$, i.e.~$M = m_H$.\footnote{We note that in the 2HDM interpretation presented 
\GW{by ATLAS}
almost the same benchmark scenario was considered~\cite{ATLAS-CONF-2023-034}.
However, therein the condition $M = m_A$ was used instead of $M = m_H$ as applied here.
We apply the latter condition in order to have a theoretically consistent form of
the Higgs potential for $m_A = m_{H^\pm} \gg m_H$, whereas the condition used by ATLAS
gives rise to an unbounded Higgs potential, and thus an unstable EW vacuum, in the parameter
space regions in which the $A \to ZH$ decay is kinematically allowed.} In addition, the alignment
limit $\cos(\alpha - \beta) = 0$ is assumed, in which the properties of the (in this case)
lightest Higgs boson~$h$ with mass $m_h = 125\gev$ 
\GW{are the same as for the}
SM Higgs boson \htb{at tree level}.
These conditions on the parameter space allow for sizable \jmn{$m_A - m_H$ mass splittings, driven by the quartic couplings in the 2HDM scalar potential~\eqref{eq:scalarpot}, facilitating the presence of a FOEWPT~\cite{Dorsch:2014qja,Biekotter:2022kgf}} 
\GW{while being in agreement with the LHC measurements of the properties of the detected Higgs boson at 125~GeV as well as with the results for the EW precision observables and further theoretical constraints.}
\GW{After imposing the above-mentioned conditions, the only remaining free parameters are}
the masses~$m_H$
and $m_A = m_{H^\pm}$ \GW{as well as} $\tan\beta$.
In the following, we \jmn{discuss} 
our results in the $(m_H,m_A)$-plane for \jmn{different values of $\tan\beta$ within the two regimes discussed above.}

\subsubsection{Low $\tan\beta$-region}
\label{sec:lowtanbeta}

In this section we present our results for the low-$\tan\beta$ regime, \jmn{which focuses on the range} 
$1 \leq \tan\beta \leq 3$. The lower bound on $\tan\beta$ was chosen because values of
$\tan\beta$ below 1 are in strong tension with constraints from flavor physics observables.
The indirect limits from flavor physics also constrain the 2HDM parameter space for slightly
larger values of $\tan\beta$ depending on the 2HDM Yukawa type and the mass of the charged \GW{Higgs boson}. \KRn{As discussed above, we do not 
carry out a detailed investigation of the
indirect limits from flavor physics in the following.}

\begin{figure}
\centering
\includegraphics[width=0.48\textwidth]{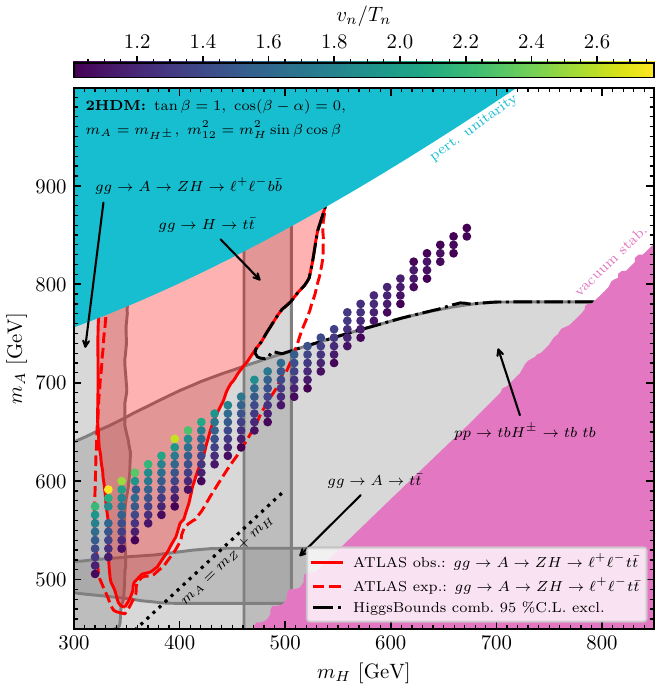}~
\includegraphics[width=0.48\textwidth]{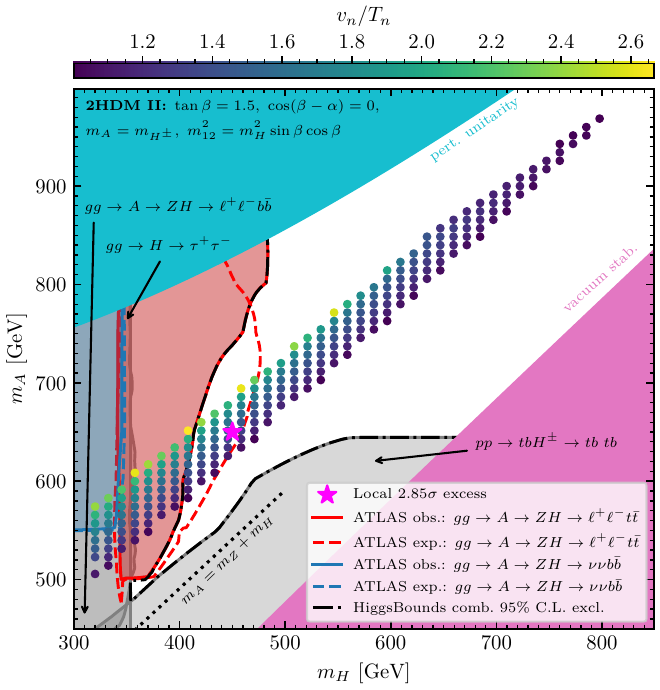}\\[0.4em]
\includegraphics[width=0.48\textwidth]{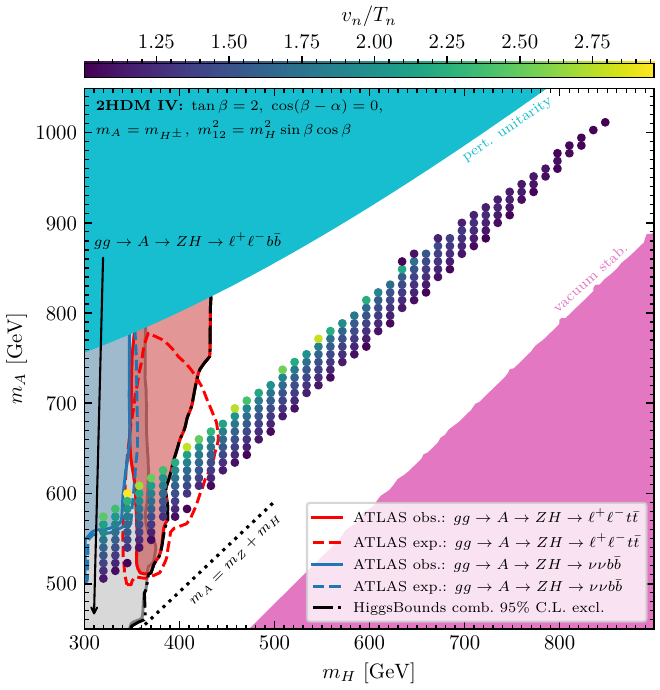}~
\includegraphics[width=0.48\textwidth]{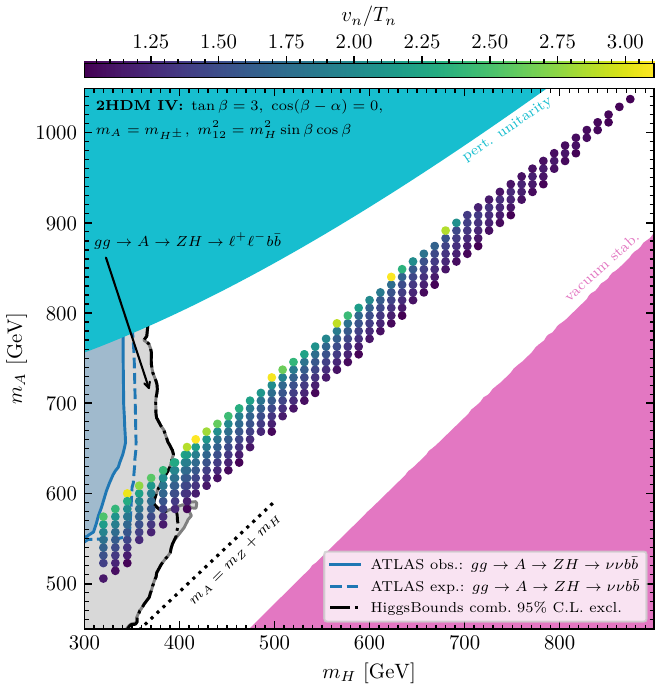}
\caption{Impact of the new ATLAS searches for the $A \to ZH$ signature in the $(m_H,m_A)$-plane for $\tan\beta = 1$
(upper left), $\tan\beta = 1.5$ and type~II (upper right), $\tan\beta = 2$ and type~IV (lower left), and
$\tan\beta = 3$ and type~IV (lower right). Parameter space regions excluded by vacuum
stability or perturbative unitarity are indicated with pink and cyan colors,
respectively. Regions excluded from previous LHC searches are indicated in gray, and regions excluded
by the new $\ell^+ \ell^- t \bar t$ and $\nu \nu b \bar b$ searches are indicated in
red and blue, respectively, where the dashed lines indicate \GW{the corresponding 
expected exclusion} 
limits. Parameter space regions featuring a FOEWPT with $v_n / T_n > 1$ are indicated
with the scatter points, where the color coding indicates the values of $v_n / T_n$.
The mass values of the most significant excess ($2.85\,\sigma$ local significance)
observed by ATLAS in the $\ell^+ \ell^- t \bar t$
search are indicated with a magenta star
in the upper right plot.}
\label{fig:lowtanbeta}
\end{figure}

In \reffi{fig:lowtanbeta} we show the impact of the \jmn{new $A \to ZH$ 
searches from ATLAS~\cite{ATLAS:2020gxx,ATLAS-CONF-2023-034}} 
in the $(m_H,m_A)$-plane for $\tan\beta = 1$ (upper left), $\tan\beta = 1.5$ (upper right),
$\tan\beta = 2$ (lower left) and $\tan\beta = 3$ (lower right).
The upper left plot is valid independently of the chosen 2HDM Yukawa type. However, for $\tan\beta \neq 1$ the
relevant cross sections and branching ratios depend on the Yukawa type, and the 
\jmn{specific choice of type \GW{that is specified in} the upper right
and the lower plots will be \GW{further} discussed below.}
%
In each plot we indicate the parts of the parameter space that are excluded by vacuum stability and perturbative
unitarity with pink and cyan colors, respectively. \jmn{The regions excluded
by the new ATLAS search~\cite{ATLAS-CONF-2023-034} for $gg \to A \to ZH$ in the 
$\ell^+ \ell^- t \bar t$ and the $\nu \nu b \bar b$ final states are indicated with 
red and blue shaded contours, \GW{respectively}, whereas}
regions excluded from 
\GW{previous}
LHC searches \jmn{(including the recent ATLAS $gg \to A \to ZH \to \ell^+ \ell^- b \bar b$ search~\cite{ATLAS:2020gxx})} are indicated in gray. 
\GW{In each case the
search channel giving rise to the exclusion (under the assumption that this 
search is applied individually, see the discussion in \refse{sec:hbimple})
is stated in the plots.}
For the new ATLAS 
\GW{searches}
we show in addition the 
\GW{expected exclusion regions} 
with dashed lines in the same colors. By comparing the gray shaded areas with the 
red and blue shaded areas, one can determine
\GW{to what extent}
the new ATLAS \GW{searches 
probe previously unexplored} 
parameter space regions.

\GW{While, as discussed in \refse{sec:hbimple}, the excluded regions resulting 
from applying each of the searches individually are shown for illustration, 
we obtain the region that is overall excluded at the 95\% C.L.}\ 
from the application of \texttt{HiggsBounds}. 
\GW{This region, which is obtained by applying for each BSM scalar only the 
observed limit of the search that has the highest expected sensitivity,} 
is indicated with the black dotted-dashed lines.
If observed limits show a 
significant excess or a significant underfluctuation in comparison to the expected 
limits, the overall limit at the 95\% C.L.\ obtained in this way 
can be weaker than the exclusion that would result from the requirement that 
each limit should be fulfilled individually. 
This feature is visible in the upper left plot for $m_A$ values slightly above 700~GeV. Here the $gg \to A \to ZH \to \ell^+ \ell^- t \bar t$ channel has the 
highest sensitivity, but since the observed limit is weaker than the expected one, a parameter region that could nominally be excluded by the $gg \to H \to t \bar t$ 
search (indicated in gray) remains unexcluded because of the adopted procedure for 
obtaining a 95\% C.L.\ exclusion.

Finally, we show in the plots in \reffi{fig:lowtanbeta} the parameter \jmn{regions} that exhibit a strong FOEWPT \jmn{as defined in  \refse{sec:thermal}} (based on the
one-loop thermal effective potential with \jmn{daisy} resummation \jmn{in} the so-called Arnold-Espinosa
scheme). 
\GW{As discussed above, we note
the sizable theoretical uncertainties in the predictions for a strong FOEWPT using this approach, and thus}
the regions shown should only be regarded as indicative for the presence of such strong transitions.
The color coding of the points indicates the ratio \jmn{between} the vev $v_n$ in the broken phase at the onset of the
transition and the nucleation temperature $T_n$. 

\medskip

\noindent $\bullet\,$ {\boldmath$\tan \beta = 1$},
\textbf{all types}

\vspace{1mm}

\noindent The results for $\tan\beta = 1$ are shown in the upper-left plot of \reffi{fig:lowtanbeta}.
One can see that the \jmn{new $A \to ZH \to \ell^+ \ell^- t \bar t$ ATLAS search (red) excludes the region  $350\gev \lesssim m_H \lesssim 450\gev$
and $650\gev \lesssim m_A = m_{H^\pm} \lesssim 800\gev$, which was so far allowed. This demonstrates} the exclusion power
of \jmn{such} smoking-gun signature for masses above the di-top threshold. \jmn{In addition, when combined}
with searches for the charged scalars using the $H^\pm \to tb$ decay~\cite{CMS:2020imj,ATLAS:2021upq},
searches for neutral scalars decaying into top-quark pairs~\cite{CMS:2019pzc}, and searches for the $A \to ZH$ decay in the $\ell^+ \ell^- b \bar b$ final
state~\cite{ATLAS:2020gxx}, \jmn{the mass range} 
$300\gev \lesssim m_H \lesssim 450\gev$ and $450\gev \lesssim m_A = m_{H^\pm} \lesssim 700\gev$ is
now excluded.
\jmn{\reffi{fig:lowtanbeta} also highlights that for $\tan\beta = 1$ the parameter region with a strong FOEWPT to which the new ATLAS search 
is sensitive, assuming $m_A = m_{H^\pm}$, 
is already excluded} 
by the charged Higgs-boson searches. 
\jmn{Yet, we stress that  
if the condition $m_A = m_{H^\pm}$ were relaxed by allowing} for an additional mass gap between these
states, i.e.~$m_{H^\pm} > m_A$ \GW{(which however would lead to a tension with the electroweak precision observables)}, the searches for the charged scalars 
\GW{would} become less sensitive, such that
the smoking gun search 
\GW{would have the highest sensitivity}
in an even larger region of parameter space.

\medskip

\noindent $\bullet\,$ {\boldmath$\tan \beta = 1.5$},
\textbf{type II}

\vspace{1mm}

\noindent The results for $\tan\beta = 1.5$ are shown in the upper right plot of \reffi{fig:lowtanbeta}. 
\jmn{While} for low $\tan\beta$ values the gluon-fusion production cross sections of~$A$ are dominantly mediated by the top-quark loop,
\jmn{making} the cross sections still \jmn{very much} independent of the type, the branching ratios of~$A$ and~$H$ differ depending on the chosen type. However, for $\tan\beta = 1.5$ the differences between
the types are mild, and we focus on the Yukawa type~II for \jmn{definiteness.} 
Comparing to the results for $\tan\beta = 1$ (upper left plot), one can see that the region excluded
by the searches for the charged scalars \jmn{via $pp \to H^\pm t b \to tb \, tb$,}
where the cross section \GW{times branching ratio} roughly scales with $1 / \tan^2\beta$ \htmm{in the low-$\tan\beta$ regime}, is substantially smaller. This search loses even more sensitivity \GW{where} the decay $H^\pm \to W^\pm H$ is kinematically allowed,
giving rise to the slope of the corresponding excluded region for $m_H \lesssim 500\gev$ \GW{(which is more pronounced than for $\tan\beta = 1$ because of the reduced $H^\pm tb$ coupling).}
As a consequence, \GW{for $\tan\beta = 1.5$} the $H^\pm \to tb$ searches are not sensitive anymore to the parameter space
region indicative \jmn{of a strong} FOEWPT. Instead, this region is excluded up to masses of
$m_H \approx 2\, m_t$ by searches for $H \to \tau^+ \tau^-$~\cite{ATLAS:2020zms,CMS:2022goy}
and by searches for the $A \to ZH$ decay using the $\ell^+ \ell^- b \bar b$ final
state~\cite{ATLAS:2020gxx}. Above the di-top threshold, the decay $H \to t \bar t$ \jmn{very quickly dominates,} and the new ATLAS \jmn{search in the $\ell^+ \ell^- t \bar t$ final state} is the most sensitive \jmn{one}. 
\GW{In contrast} to the $\tan\beta = 1$ case,
for $\tan\beta = 1.5$ the new search \jmn{is able to exclude a significant parameter region featuring a strong FOEWPT that was previously allowed.} \GW{The new search substantially pushes} the lower limit on the
Higgs boson masses to larger values of about $m_H \gtrsim 400\gev$ and
$m_A = m_{H^\pm} \gtrsim \GW{550}\gev$.
\jmn{We also stress that,} based on the expected cross section limits, \jmn{an even larger mass region} would be excluded, 
as indicated with the dashed red line. However, ATLAS observed a local $2.85\sigma$ excess 
\GW{for $m_A \approx 650\gev$ and $m_H \approx 450\gev$}, giving rise to a weaker observed cross section limit. 
\jmn{The} masses corresponding to the excess, indicated with a magenta star in the upper right
plot of \reffi{fig:lowtanbeta}, and the corresponding cross section are such that they fall into the \jmn{strong FOEWPT region.} 
In \refse{sec:excess} we will discuss in \jmn{greater detail the tantalizing possibility of such an excess to be the first experimental hint of a strong FOEWPT within} 
the 2HDM. We will give a broad characterization of the FOEWPT predicted by
this benchmark scenario, focusing on whether the scenario might be suitable for a realization of
EW baryogenesis, and whether the associated GW signal
might be detectable with LISA.

\GW{As an important outcome of the above discussion,} 
a promising complementary LHC search to target the \jmn{strong FOEWPT} region consists of \jmn{charged scalar production followed by the decay $H^\pm \to W^\pm H \to \ell^\pm \nu t \bar t$,} 
which so far has not been performed.\footnote{Searches targeting the $H^\pm \to W^\pm H$ decay have been performed by CMS assuming the decay $H \to \tau^+ \tau^-$ and assuming a fixed mass of $m_H = 200\gev$~\cite{CMS:2022jqc}.}
\jmn{In particular, producing the charged scalar via $p p \to t b H^\pm$ would in this case lead to 
a 4-top-like (or 3-top-like, depending on the signal selection) signature, which has very recently been performed by CMS~\cite{CMS:2023ftu} and ATLAS~\cite{ATLAS:2023ajo} (but not interpreted in terms of the scenario discussed here), yielding a mild excess over the SM expectation.}

Finally, it can be seen that for $\tan\beta = 1.5$
the new smoking gun search using the $\nu \nu b \bar b$
final state
starts to probe the considered parameter plane.
An exclusion region is visible below the di-top
threshold regarding $m_H$ and 
\GW{for}
a minimum amount of
mass splitting of $m_A - m_H \gtrsim 200\gev$.
However, in contrast to the searches using the
$\ell^+ \ell^- t \bar t$ final state
indicated by the red shaded region, the blue shaded
region indicating the new exclusion region
resulting from the
search using the $\nu \nu b \bar b$ final state
is already excluded
by previous LHC searches, namely searches for
$H$ decaying into tau-lepton
pairs~\cite{ATLAS:2020zms,CMS:2022goy} and searches
for the smoking gun signature $A \to ZH$ with
\GW{$Z \to \ell^+ \ell^-$ and the}
decay of $H$ into bottom-quark
pairs~\cite{ATLAS:2020gxx}. One should note, however,
\htmm{that the new $A \to ZH$ 
search in the $\nu \nu b \bar b$ final state covers
larger masses up to $m_H = 600\gev$ and
$m_A = 1000\gev$~\cite{ATLAS-CONF-2023-034},
extending the reach of previous ATLAS searches
in $\ell^+ \ell^- b \bar b$ and
$\ell^+ \ell^- W^+ W^-$ final states~\cite{ATLAS:2020gxx} in the
region with $m_H > 350\gev$ and $m_A > 800\gev$.}
\htmm{In the 2HDM constraints from perturbative
unitarity
(cyan area in \reffi{fig:lowtanbeta}) exclude
large mass splittings between states from the
same SU(2) doublet. As a consequence,
the extended mass reach of the new searches
in the $\nu \nu b \bar b$ final state
\GW{(not visible in the plot)}
does not give rise to new constraints
on the 2HDM
for $m_A > 800\gev$.
However, in other models allowing for larger
mass splittings between the BSM states, the
searches in the $\nu \nu b \bar b$ final state
can potentially provide new constraints}.


\medskip

\noindent $\bullet\,$ {\boldmath$\tan \beta = 2$},
\textbf{type IV}

\vspace{1mm}

\noindent We show the results for $\tan\beta = 2$
in the lower \KRn{left} plot
of \reffi{fig:lowtanbeta}. 
From here on, we focus our discussion on
the Yukawa type~IV, in which the new ATLAS
searches have the 
\GW{highest potential for probing parameter regions that were
unconstrained so far}.
In particular, compared to type~I and~III
the decay width for $H \to b \bar b$ is enhanced
in type~IV for $\tan\beta > 1$, such
that the searches in the $\nu \nu b \bar b$
final state become more important with increasing
values of~$\tan\beta$. Moreover,
in type~IV the decay width
for $H \to \tau^+ \tau^-$ is suppressed \jmn{approximately}
by $1 / \tan^2 \beta$,
whereas it is enhanced by \GW{about a factor}
of~$\tan^2\beta$ in type~II. 
\GW{Hence, while in type~II the parameter region below the
di-top threshold, i.e.~$m_H < 2 m_t$, is
entirely excluded by the searches for
di-tau resonances, in type~IV}
\htb{the $\nu \nu b \bar b$ search \GW{can potentially} 
yield stronger constraints.}

\GW{One can see} in the lower 
\GW{left}
plot of \reffi{fig:lowtanbeta}
\GW{that in this case only three
LHC searches 
give rise to excluded regions in}
the parameter plane. This is a manifestation
of the fact that 
\GW{the so-called wedge-region} of the
2HDM, with intermediate
values of $2 \lesssim \tan\beta \lesssim 8$, 
is difficult to probe at the LHC~\cite{Gori:2016zto}.
As an example, we note that
the searches for the charged
scalars via the signature
$pp \to H^\pm t b \to tb \, tb$,
suppressed by factors of \GW{about}~$1 / \tan^2\beta$
\htmm{in the low-$\tan\beta$ regime},
cannot probe 
the parameter plane 
\GW{in this case}.
Below the di-top threshold, we find that
the $A \to ZH$ searches in the
$\ell^+ \ell^- b \bar b$ (gray) and
the $\nu \nu b \bar b$ (blue) final states
exclude the entire
region allowed by the theoretical constraints.
As discussed above, \htmm{for 
$m_A < 800~$GeV searches for the decay $A \to ZH$
using the decay $Z \to \ell^+ \ell^-$
have been performed by ATLAS~\cite{ATLAS:2020gxx},
\GW{which} are more powerful
than the new searches using the $Z \to \nu \nu$ decay
(the corresponding CMS search using
the $Z \to \ell^+ \ell^-$ decay covers masses
up to $m_A = 1$~TeV, but is based on first-year
Run~2 data only~\cite{Sirunyan:2019wrn}).
For $m_A > 800$~GeV ATLAS limits exist
only from the new searches using
the decay $Z \to \nu \nu$
(the resulting exclusion regions
are not visible in our plots since in the 2HDM such large mass splittings
are excluded by perturbative unitarity, \GW{indicated by the} cyan area).}
Above the di-top threshold, the searches
relying on the decay $H \to b \bar b$ quickly
lose their sensitivity to the 2HDM parameter plane.
Accordingly, for masses of~$H$ substantially larger
than twice the top-quark mass
the new smoking gun search 
\GW{for}
the decay $H \to t \bar t$ 
\GW{is in fact} the only channel
that can probe the parameter plane.
As indicated with the red shaded area,
the searches in the $\ell^+ \ell^- t \bar t$ final state
are able \KRn{to exclude masses smaller}
than $m_H \approx 400\gev$
and $m_A \approx 750\gev$ for the lighter and
the heavier BSM resonance, respectively.
As \GW{it is also visible in the plots 
for $\tan\beta = 1$, $\tan\beta = 1.5$ and 
$\tan\beta = 2$}
of \reffi{fig:lowtanbeta}, the difference between
\GW{the} expected (red dashed line)
and \GW{the} observed (red solid line) exclusion region
resulting from the searches using the
$\ell^+ \ell^- t \bar t$ final state 
\GW{arises}
from the excess observed \GW{in the ATLAS 
search (except for the upper right part of the red region 
in the plots for $\tan\beta = 1.5$ and 
$\tan\beta = 2$, where the observed limit is stronger than the expected one).}

\medskip

\noindent $\bullet\,$ {\boldmath$\tan \beta = 3$},
\textbf{type IV}

\vspace{1mm}

\noindent \GW{As a final step of}
the discussion of the low-$\tan\beta$
regime \GW{we consider} a value of~$\tan\beta = 3$.
The results of our analysis are shown in the
lower right plot of \reffi{fig:lowtanbeta}.
Again, we focus on the Yukawa type~IV (see the
discussion above).
One can see that in this case the smoking
gun searches in the $\ell^+ \ell^- t \bar t$
final state cannot probe the parameter space
\GW{as a consequence of the suppression of}
the gluon-fusion production
cross section of~$A$.
We will discuss in \refse{sec:prospects} the prospects
for probing the benchmark plane for
$\tan\beta = 3$ in future runs of the LHC,
in which roughly 20 times more integrated
luminosity will be collected by both
ATLAS and CMS.\footnote{See
\citere{Biekotter:2022kgf}
for an earlier projection based on
expected cross section limits reported
by CMS.}
At and below the di-top threshold
$m_H \approx 2 m_t$ the results are similar to the
case of $\tan\beta = 2$, where the smoking
gun searches relying on the decay $H \to b \bar b$
\GW{essentially exclude the whole parameter region}.
One should note that in type~IV (and type~II)
the partial widths for the decays
$A,H \to t \bar t$ are suppressed \jmn{approximately} by
$1 / \tan^2\beta$, and 
the partial width for the decay
$H \to b \bar b$ is \jmn{conversely} enhanced
by \jmn{(approximately)} $\tan^2\beta$.
As a result, the gray exclusion region from the
searches in the
$A \to ZH \to \ell^+ \ell^- b \bar b$ channel
\GW{extends to}
slightly larger masses
for $\tan\beta = 3$ compared to
$\tan\beta = 2$ (lower left plot).

\subsubsection{High $\tan\beta$-region}
\label{sec:hightanbeta}
In the discussion above, we investigated the
low-$\tan\beta$ regime in which the 
\GW{CP-odd Higgs boson} $A$
can be produced with a sizable cross section via
gluon-fusion. On the other hand,
for large values of $\tan\beta \gtrsim 10$ the
gluon-fusion production mode is suppressed.
In the Yukawa types~II and~IV of the 2HDM,
$A$ is then produced more efficiently
via $b \bar b$-associated
production, which is enhanced by \jmn{about}~$\tan^2\beta$ in these types.
Consequently, focusing
on the high-$\tan\beta$ regime here,
we expect the searches for the signature
$A \to ZH$ assuming $b \bar b$-associated production
to become relevant in type~II and type~IV.
Since in type~IV the limits from searches for
scalar resonances decaying into tau-lepton pairs
are substantially weaker (see the discussion above),
we investigate here the impact of the new ATLAS
searches on the type~IV 2HDM parameter space.

It should be noted that
the new ATLAS searches reported
in \citere{ATLAS-CONF-2023-034}
only considered the $b \bar b$-associated
production utilizing the $\nu \nu b \bar b$
final state, whereas the smoking-gun search
utilizing the $\ell^+ \ell^- t \bar t$ final state
was considered only assuming gluon-fusion
production of the heavy BSM resonance.
Thus, the only relevant searches for the $A \to ZH$ decay
in the following discussion will be
the previously reported searches utilizing
the $\ell^+ \ell^- b \bar b$ final
state~\cite{ATLAS:2020gxx,Khachatryan:2016are}
and the new searches utilizing the $\nu \nu b \bar b$
final state~\cite{ATLAS-CONF-2023-034}.

\begin{figure}
\centering
\includegraphics[width=0.55\textwidth]{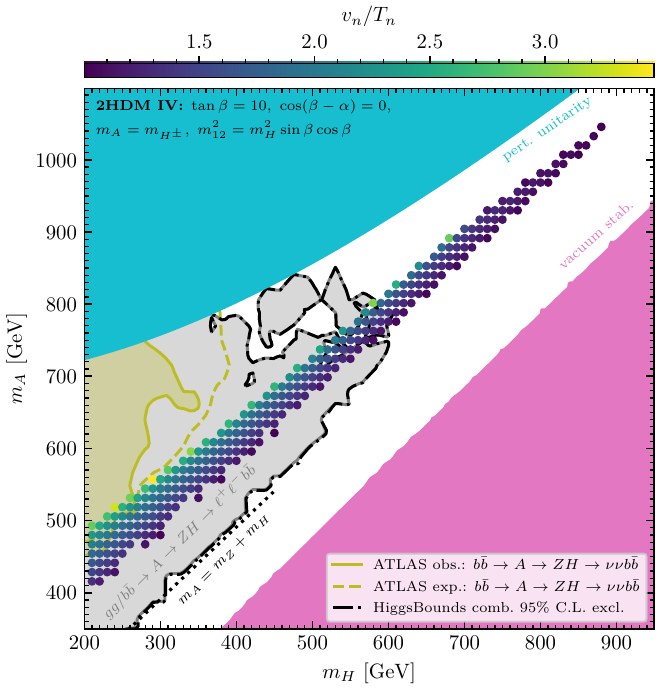}
\caption{As in \reffi{fig:lowtanbeta}, but
for $\tan\beta = 10$ in type~IV. Parameter space
regions excluded by the new $\nu\nu b \bar b$
searches in the $b \bar b$-associated production channel
are indicated in yellow, 
\GW{while} the yellow dashed
line \GW{indicates the expected exclusion limit}.} 
\label{fig:hightanbeta}
\end{figure}

In \reffi{fig:hightanbeta} we show our results for
$\tan\beta = 10$ as a representative benchmark
scenario for the high-$\tan\beta$ regime.
The color coding of the exclusion regions and the
scatter points is the same as in \reffi{fig:lowtanbeta},
except for the yellow dashed and solid lines
indicating the expected and observed
exclusion limits resulting from the
recent ATLAS search for
$b \bar b \to A \to ZH \to \nu \nu b \bar b$,
respectively. One can see that the parameter space
region excluded by this search (yellow shaded area)
lies within the gray shaded area indicating the
exclusion from the searches for $b \bar b
\to A \to ZH
\to \ell^+ \ell^- b \bar b$~\cite{ATLAS:2020gxx},
which were published previously.
Hence, although the new searches 
\GW{based on} the decay of
the $Z$-boson into neutrinos are able to probe the
2HDM parameter space for values of $\tan\beta \gtrsim 10$,
these regions are already excluded by the searches
\GW{making use of}
the decay of the $Z$-boson into charged leptons.
We stress, however, that
the new searches using the $\nu \nu b \bar b$
final state cover a larger mass interval of up to
1.2~TeV for the heavy BSM resonance 
\GW{(not visible in the plot)}, whereas 
the
corresponding upper limit in the ATLAS
searches using the $\ell^+ \ell^- b \bar b$ final state
is \GW{about} $800\gev$. Therefore, in other models in which
larger mass \GW{splittings} between the heavier and the
lighter BSM resonance \hto{are} possible compared to the
2HDM (where perturbative unitarity implies an
upper limit on such mass splittings, see the cyan
region in \reffi{fig:hightanbeta}), the new searches
using the $\nu \nu b \bar b$
final state could potentially give
rise to new constraints.

The two LHC searches relevant
in \reffi{fig:hightanbeta} differ in
the 
\GW{targeted decay mode} of the
$Z$ boson, whose branching ratios are precisely
measured. As a consequence, the relative importance
of both searches \GW{is} independent of the 2HDM
parameters, especially of $\tan\beta$. We can therefore
extrapolate based on the results
for $\tan\beta = 10$ shown in
\reffi{fig:hightanbeta} that also for larger values
of $\tan\beta$ the searches \GW{making use of} the
$Z \to \ell^+ \ell^-$ decay mode are more promising
to probe the considered benchmark plane compared
to the searches using the $Z \to \nu \nu$ decay mode.
It should also be taken into account that for larger
values of $\tan\beta$ other LHC searches become
relevant in type~IV.\footnote{In type~II,
for $\tan\beta \gtrsim 10$ the
whole investigated
parameter plane is excluded for masses
up to about 1~TeV by searches for scalar
resonances decaying into tau-lepton
pairs~\cite{ATLAS:2020zms,CMS:2022goy,CMS-PAS-EXO-21-018}.}
In particular, searches
for new resonances produced in $b \bar b$-associated
production with subsequent decay into bottom-quark
pairs~\cite{ATLAS:2019tpq},
giving rise to four $b$-jet final states,
start to exclude sizable parts of the benchmark plane
for $\tan\beta 
\gtrsim 15$.
Moreover, for such values of $\tan\beta$ searches
for new resonances
\htmm{produced in association with a photon 
and
decaying into two jets~\cite{ATLAS:2019itm}}
are able
to exclude parameter regions
especially in the mass-degenerate
regime.

\subsection{Future prospects for $\ell^+ \ell^- t \bar t$ searches}
\label{sec:prospects}

In \refse{sec:lowtanbeta} we have demonstrated that
the new ATLAS smoking-gun searches 
targeting the $\ell^+ \ell^- t \bar t$
final state exclude sizable parts
of previously allowed parameter space of
the 2HDM assuming values of $\tan\beta$ not much
larger than one. In particular, we have shown that
for BSM scalar masses above the di-top threshold and
values of $\htb{1.5} \lesssim \tan\beta \lesssim 3$ the smoking-gun
searches 
\jmn{arguably are}
the most promising of all LHC searches
\GW{for probing so far unexplored} parameter space regions,
\GW{with the potential}
to discover additional
Higgs bosons that are consistent with a 2HDM
interpretation. Due to their exceptional importance,
we briefly discuss here the projected sensitivity
of the searches for the $A \to ZH$ decay 
\GW{in the $\ell^+ \ell^- t \bar t$ final state}
during future runs of the LHC and the
High-Luminosity LHC~(HL-LHC). 
\GW{As input for our projections we use the expected limits from the ATLAS 
analysis for an integrated luminosity of \htmm{140}~fb$^{-1}$. This improves} 
upon the previous projections presented in
\citere{Biekotter:2022kgf} that \GW{were} obtained based
on \GW{an estimate of the} expected sensitivities from the CMS
collaboration.

\GW{In \refap{app:proj} we provide a comparison of the two projections, showing that they are in good agreement with each other in view of the 
systematic uncertainties of the analyses.}

The projected exclusion limits discussed
in the following were obtained by
re-scaling the expected cross-section
limits reported in \citere{ATLAS-CONF-2023-034}
with future values for the integrated luminosity that
will be collected during future runs of
the (HL-)LHC, i.e.\
\begin{equation}
\sigma^{\mathrm{exp.}~95\%~\mathrm{CL}}_{\rm proj.}(\mathcal{L},m_H,m_A) =
\sigma^{\mathrm{exp.}~95\%~\mathrm{CL}}_{\rm Run~2}(m_H,m_A)
\; \sqrt{\frac{140~\mathrm{fb}^{-1}}
{\mathcal{L}}} \ .
\end{equation}
Here, $\sigma^{\mathrm{exp.}~95\%~\mathrm{CL}}_{\rm Run~2}$
is the expected cross-section
limit at 95\% confidence level
reported by ATLAS based on 140~fb$^{-1}$
collected during Run~2 as a function
of the masses of the \GW{probed BSM resonances}, 
and $\sigma^{\mathrm{exp.}~95\%~\mathrm{CL}}_{\rm proj.}$
is the future projection of the expected
cross-section limits depending additionally
on the assumed integrated luminosity $\mathcal{L}$.
Accordingly, in the projections we only account
for the reduction of statistical uncertainties,
whereas no assumption is made on
possible improvements of systematic theoretical
or experimental uncertainties.
Moreover, we do not account for the slight
increase of the center-of-mass energy at future
runs of the LHC and the HL-LHC, operating at 13.6~TeV 
and 14~TeV, respectively, compared to the Run~2
dataset collected at 13~TeV.
Taking this into account, we consider our projections
as fairly conservative estimates.

\begin{figure}
\centering
\includegraphics[width=0.48\textwidth]{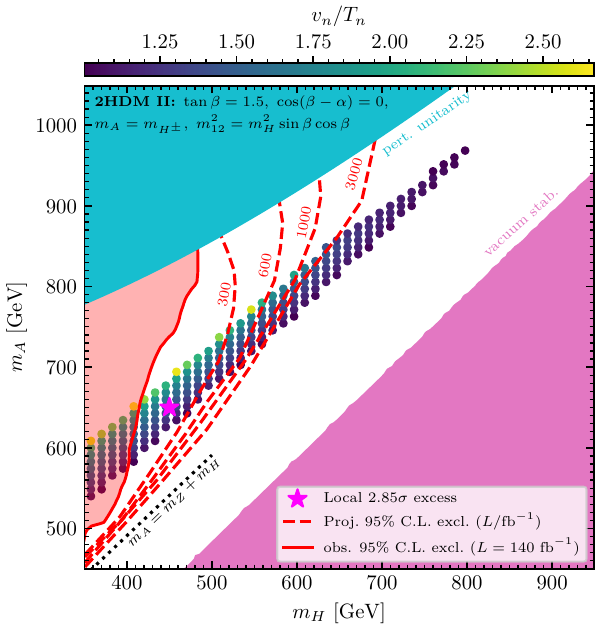}~
\includegraphics[width=0.48\textwidth]{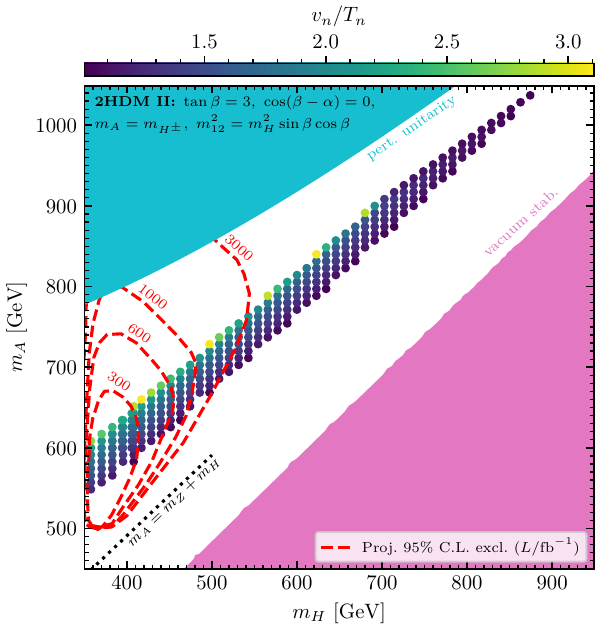}
\caption{As in \reffi{fig:lowtanbeta}
for $\tan\beta = 1.5$ (left) and
$\tan\beta = 3$ (right), shown here for type~II, 
but
the red dashed lines indicate projected
\GW{expected}
exclusion regions assuming integrated luminosities
of $300,600,1000,3000~\mathrm{fb}^{-1}$ 
\GW{from} future runs of the LHC.}
\label{fig:proj}
\end{figure}

The projected expected cross section limits
can be cast into projected exclusion regions
in the 2HDM.
In \reffi{fig:proj} we show our projections
in the 2HDM benchmark plane introduced in
\refse{sec:newconstr} 
\GW{for the Yukawa type~II with}
$\tan\beta = 1.5$ in the left
plot and $\tan\beta = 3$ in the right plot.
In both plots,
the color coding of the scatter points and the
definition of the pink and cyan
regions is as in \reffi{fig:lowtanbeta},
and the red dashed lines indicate the \GW{expected} exclusion
regions for different values of the integrated
luminosity, ranging from $\mathcal{L} = 300~\mathrm{fb}^{-1}$
(end of LHC Run~3) to $\mathcal{L} = 3000~\mathrm{fb}^{-1}$
(end of the LHC high-luminosity phase).
Moreover, in the left plot the red shaded area
indicates the currently excluded region based on
the observed cross section limits 
\GW{obtained for}
$\mathcal{L} = \htmm{140}~\mathrm{fb}^{-1}$, and the
magenta star indicates the masses for which
ATLAS has observed the most pronounced local
excess (see \refse{sec:excess}).
As already discussed in
\refse{sec:lowtanbeta},
currently the smoking-gun searches are not
able to probe the benchmark plane for
$\tan\beta = 3$
(see the lower left plot of \reffi{fig:lowtanbeta}).
Accordingly, no red shaded region is visible
in the right plot of \reffi{fig:proj}.

One can observe in the left plot of
\reffi{fig:proj} that 
\GW{with the prospective improvements of the integrated luminosity 
it will be possible to increase very significantly the regions that 
can be probed}
in the considered benchmark plane 
\GW{for $\tan\beta = 1.5$}.
\GW{While currently in the upper right part of the red shaded region} 
the smoking-gun searches are
able to exclude masses \GW{up to values slightly below 500~GeV   
for the lighter and up to} 850~GeV for the heavier
BSM scalar, in the future the LHC will be able to
probe \GW{via this search} masses up to about 700~GeV and 1~TeV
for the lighter and the heavier BSM scalar,
respectively. 
\GW{This improvement in sensitivity has a very important impact 
on the parameter region that is} suitable for the realization of a strong 
FOEWPT according to the
thermal effective potential approach \htb{(as described in \refse{sec:thermal})}.
\GW{In the case of the absence of a signal the exclusion within 
the region that is indicative for a strong FOEWPT would extend up to}
$m_H \lesssim 550\gev$ and
$m_A \lesssim 700\gev$.
\GW{It should be noted in this context}
that the
strength of the phase
transition diminishes with
increasing masses of the BSM scalars. 
\GW{As one can infer from the 
color coding of the displayed points,}
the projected exclusion regions
cover the parameter \GW{region} for which the strongest
phase transitions can be accommodated.
\GW{As a result,
\htmm{and since in the 2HDM the generation of a
sufficiently large BAU may be possible
only for small values of $\tan\beta$
not much larger than
one~\cite{Fromme:2006cm},}
the searches for the smoking-gun signature will provide 
a stringent test of} 
the possibility to explain the~BAU
by means of EW baryogenesis in the 2HDM.

In this context it is also important to note
that in the 2HDM
the primordial GW background generated during
the phase transition 
\GW{is only potentially detectable
with LISA for the largest possible values of~$v_n / T_n$, which 
are only reached in a very restricted region of the 2HDM parameter space 
and have a very strong dependence on the details of the 
scalar spectrum}~\cite{Biekotter:2022kgf}.
We have verified using the approach
discussed in \refse{sec:gw}
that \htmm{for the considered values
of $\tan\beta$} all parameter points
predicting a GW signal \GW{that is potentially} detectable with LISA
would be 
\GW{probed}
by the projected exclusion
limits from the HL-LHC.
Hence, in the 2HDM
the HL-LHC results will have an enormous
impact on the possibility for a detection of
a GW background with LISA
consistent with a FOEWPT.
This exemplifies that the HL-LHC has the
potential to
\GW{probe} large
parts of the relevant parameter space before
the LISA experiment will have started its operation.
\htmm{Here it should be noted, however,
that
the presence of a strong FOEWPT, without
demanding a realization of EW baryogenesis,
is also possible for larger values
of~$\tan\beta$,
where the $gg \to A \to ZH \to
\ell^+ \ell^- t \bar t$ searches
lose their sensitivity.
A GW signal potentially detectable with LISA
therefore \GW{cannot} be fully probed with
the searches in the $\ell^+ \ell^- t \bar t$
final state.}

\GW{Besides the analysis of the potential of future runs of the (HL-)LHC
for probing the 2HDM parameter space in terms of projected exclusion 
limits, it is also of interest to investigate the possible}  
interplay between
the LHC and LISA 
\hto{for the case where}
a smoking-gun signal \GW{would be detected}.
\GW{We note in this context that}
the magenta star indicating the
mass values corresponding to
the excess observed by ATLAS lies well within
the \jmn{discovery} reach of the LHC, quite possibly already after
the end of Run~3. The detection of the smoking-gun
signal would allow for the determination
of~$m_H$ and $m_A$, and possibly also of
$m_{H^\pm}$ via the corresponding cross sections
in combination with the application of other
constraints. The 
\GW{experimentally determined}
values of the BSM scalar masses
could then be used in \jmn{dedicated} analyses
of the phase transition dynamics.  
For instance, the 
\GW{experimental information about the mass hierarchy of the}
scalar spectrum would allow \GW{an} analysis
of the thermal potential in \GW{an appropriately chosen}
dimensionally-reduced effective-field theory,
in which the heavier scalars \GW{have been} integrated
out in a systematic way in order to facilitate 
\GW{the incorporation of relevant}
higher-order effects, \jmn{as well as \GW{dedicated}  
lattice simulations}
(see~\citeres{Kainulainen:2019kyp,Ekstedt:2022bff}
for recent efforts towards these
directions in the 2HDM, \jmn{and~\citeres{Niemi:2020hto,Gould:2021oba,Schicho:2021gca,Schicho:2022wty,Gould:2023ovu} for related investigations in other extended scalar sectors}).

In the right plot of \reffi{fig:proj}, in which
we show the projections for $\tan\beta = 3$,
one can see that with more integrated luminosity
the (HL)-LHC \GW{also in this case} is able to probe substantial
parts of the otherwise unconstrained parameter
space regions. Interestingly,
\htbr{the red
dashed lines indicating the expected
reach of the LHC stretch out to
the largest values
of~$m_H$} \GW{within} the parameter regions
which might be suitable for a realization
of a strong FOEWPT.
Assuming an integrated luminosity of~$3000~\mathrm{fb}^{-1}$
\GW{collected
by both} ATLAS and CMS \GW{by}
the end of the LHC high-luminosity phase, 
masses of up to $m_H \approx \GW{550}\gev$ and
$m_A \approx 800\gev$ can be probed.
Here it should be taken into account that the
parameter space region with $m_H$ below the
di-top threshold is already excluded
\htbr{by di-tau searches (only for type~II) and}
by searches
for $gg \to A \to ZH \to \ell^+ \ell^- b \bar b$ (both for type~II and type~IV),
as was discussed in detail in \refse{sec:lowtanbeta}
(see the lower right plot of \reffi{fig:lowtanbeta}).
However, the sensitivity of these searches to
the parameter space regions above the di-top
threshold will not improve significantly with
increasing data, because the branching ratio
for the decay $H \to b \bar b$ is \hto{strongly} suppressed for $m_H > 2 m_t$.

In summary, the fact that the smoking-gun search
for $gg \to A \to ZH \to \ell^+ \ell^- t \bar t$
\GW{will be} able to probe masses of $m_H > 2 m_t$
\GW{in the low $\tan\beta$ regime}
in the future is crucial for testing the 2HDM
parameter space regions suitable for an explanation of the BAU via EW baryogenesis.

\subsection{A hint of a strong 1$^{\rm st}$-order EW phase transition in the 2HDM?
}
\label{sec:excess}
\hto{We now turn to the analysis}
of the local $2.85\,\sigma$
excess observed by ATLAS. 
\GW{We investigate to what extent a scenario with the mass values}
of $m_A = 650\gev$ and $m_H = 450\gev$ \hto{can be accommodated}
in the 2HDM 
\hto{and how this scenario can be
\jmn{further} tested in the 
future.}\footnote{\TBn{While here we consider
the possibility of new physics being
the origin of the observed excess,
we note that an excess in the $t \bar t Z$
final state falls within
a class of experimental discrepancies
with respect to the SM predictions
observed at the LHC in multi-lepton 
$t \bar t + X$
final states, e.g.~for
$t \bar t W$~\cite{CMS:2022tkv,ATLAS-CONF-2023-019},
$t \bar t h$~\cite{ATLAS:2021qou,
CMS-PAS-HIG-19-011}
and
$t \bar t t \bar t$~\cite{CMS:2023ftu,
ATLAS:2023ajo} production.
The possibility of a mismodeling
of the SM expectation in multi-lepton
$t \bar t + X$ final states should also be investigated in this context.}}
To this end, we first
determine the cross section 
\GW{that would be associated with}
the excess. Since \GW{information on} the likelihoods has not
been made public by ATLAS, we settle here for
\hto{an} approximation based on the reported
95\%~C.L.\ cross-section limits.
For the mass hypothesis stated above,
ATLAS found an expected and observed limit of
0.299~pb and 0.762~pb, respectively. The
best-fit signal cross section can be estimated
in Gaussian approximation as the difference
between observed and expected limit. Furthermore,
again assuming a Gaussian distribution of the
underlying likelihood, we can determine a symmetric
uncertainty of the cross sections in such a way that
the background-only hypothesis 
deviates by the observed local significance from
the central value.
In this way we obtain
a cross section of
\begin{equation}
\sigma(gg \to A) \times \mathrm{BR}(A \to ZH)
  \times \mathrm{BR}(H \to t \bar t) =
    0.46 \pm 0.16~\mathrm{pb} \ ,
\label{eq:excess}
\end{equation}
corresponding to the excess observed by
ATLAS at the above-mentioned mass
values for 
\GW{the two types of Higgs bosons}.\footnote{Assuming a Gaussian likelihood,
another way of determining the uncertainty is
by dividing the expected limit by 1.64, which
would result in an uncertainty of
$0.18~\mathrm{pb}$,
which is in good agreement with the value given
in \refeq{eq:excess} considering the \GW{relatively} low
significance of the excess and the resulting
size of the cross-section uncertainty.}

In order to 
\GW{investigate an interpretation}
of the
excess within the 2HDM, we utilize the same
benchmark scenario as before, but now
fixing $m_H = 450\gev$ and
$m_A = m_{H^\pm} = 650\gev$.
\GW{While we adopt the mass values for which the excess observed 
by ATLAS is most pronounced, we note}
that the
mass resolution of the ATLAS search is rather \hto{coarse}.  
\GW{Thus, the excess would also be compatible with mass values in the 
vicinity of the specified values, and the}
overall
conclusions regarding a description of the excess
in the 2HDM 
\GW{would be unchanged in this case.
On the other hand,}
it is known that in the 2HDM the
GW signals produced in FOEWPTs are very sensitive
to the 
\GW{details of the spectrum of the}
scalar masses~\cite{Biekotter:2022kgf}. 
In the discussion of the GW signals of this
benchmark point and the 
\GW{analysis} whether a \GW{potentially}
detectable GW signal \jmn{would be predicted based on the excess,} 
we will therefore vary the masses of~$A$ and~$H$
in a mass window \GW{of $\pm\, 50$~GeV}
which we 
\GW{use as a rough estimate for the potential signal region}.

\begin{figure}
\centering
\includegraphics[width=0.48\textwidth]{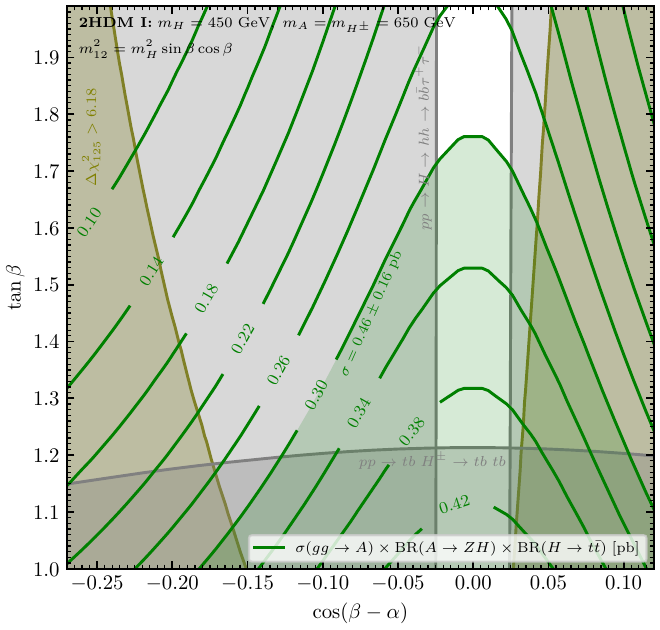}~
\includegraphics[width=0.48\textwidth]{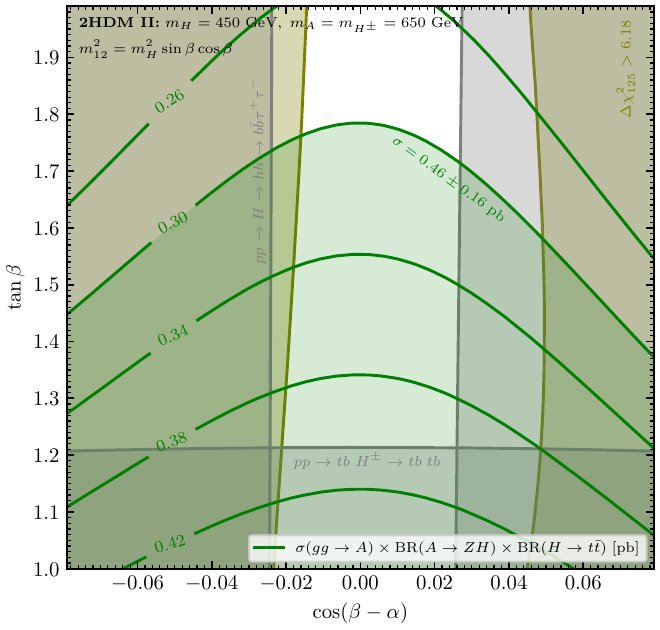}
\caption{\GW{For a description of the excess observed in the ATLAS search
within the 2HDM the}
green shaded regions are preferred
at the level of $1\sigma$ in type~I (left) and
type~II (right). \GW{The} olive shaded regions are
disfavored by the cross-section measurements
of the 125~GeV Higgs boson by more than
$2\,\sigma$ compared to the SM.
\GW{The} grey shaded regions are excluded by
LHC cross-section limits from searches
for charged Higgs bosons \TBa{and from searches
for resonant Higgs-boson pair-production
using the $H \to hh$ decay in
$b \bar b \tau^+ \tau^-$ final states}.}
\label{fig:excess}
\end{figure}

\subsubsection{Preferred parameter regions}

\GW{Accommodating the observed excess within the 2HDM implies also 
constraints on the other 2HDM parameters (besides the BSM scalar masses).
It follows}
from the
discussion in \refse{sec:newconstr} that
values of $\tan\beta \approx 1.5$ are required
for obtaining sufficiently large cross sections
to describe the excess (see the upper right plot
of \reffi{fig:lowtanbeta}). We therefore vary $\tan\beta$
in the interval $1 \leq \tan\beta \leq 2$,
and in addition we consider deviations
from the alignment limit in terms of the
free parameter $\cos(\beta - \alpha)$, see the
discussion in \refse{sec:2hdm}. 
\GW{Since we only consider}
$\tan\beta$ values \GW{close to $\tan\beta = 1$}, the theoretical 
predictions for the cross section depend only
marginally on the Yukawa type.
However, it should be
taken into account that the allowed ranges
of $\cos(\beta - \alpha)$, which is constrained
\TBa{(among others)}
by the cross-section measurements of the
Higgs boson at 125~GeV, 
\TBa{can be} different
in the different types.
We restrict the discussion here
to
\hto{type~I and~II as representative examples.}

In \reffi{fig:excess} we show the parameter
plane with $\cos(\beta - \alpha)$ on the
horizontal axis and $\tan\beta$ on the
vertical axis and the additional
2HDM parameters set according to the
discussion above, for the 2HDM type~I in the
left and for type~II in the right plot.
The olive colored regions are disfavored based on
the LHC cross-section measurements of the
Higgs boson $h$ at 125~GeV, 
where we used
\texttt{HiggsSignals}~\cite{Bechtle:2013xfa,
Bechtle:2014ewa,Bechtle:2020uwn}
(incorporated in
\texttt{HiggsTools}~\cite{Bahl:2022igd})
to perform a $\chi^2$-fit
to the various measurements.
\htg{Specifically, we demand that the $\chi^2_{125}$ value 
of a given 2HDM parameter point 
\GW{arising from the measured properties of the detected Higgs boson at 
about $125\gev$
has to be} less than $2\,\sigma$ ($\Delta\chi_{\jmn{125}}^2 \GW{<}$ 6.18) 
away from the SM \GW{result} ($\chi^2_{125,\, \rm SM} = 117.7$).}
\GW{Since up to now}
the LHC measurements regarding the \GW{properties of the 
Higgs boson at} 125~GeV
are in agreement with
the SM predictions,
the allowed 2HDM parameter region 
\GW{is located}
around $\cos(\beta - \alpha) = 0$.
The gr\hto{a}y regions
\GW{in the two plots are} excluded \KRn{by the cross section limits from LHC searches for new Higgs bosons.}
\TBa{Specifically, we find that the 
exclusion regions in the plot arise from two LHC searches. 
The gray regions at $\tan\beta \lesssim 1.2$
(both plots) are excluded
by the cross section limits from searches
for charged Higgs bosons using the $H^\pm \to tb$
decay~\cite{CMS:2020imj,ATLAS:2021upq}.
The gray regions 
at $|\cos(\beta - \alpha)| \gtrsim 0.025$
(both plots)
are excluded by searches for resonant
Higgs-boson pair production using the
$H \to hh$ decay in the $b \bar b \tau^+ \tau^-$
final state~\cite{CMS:2021yci,ATLAS:2022xzm}.
It should be noted here that searches
for resonant pair production of
the Higgs boson at~125~GeV
cannot probe the alignment limit of $\cos(\beta - \alpha) = 0$,
since
the triple scalar \KRn{$hhH$} coupling vanishes
at tree level in this limit.}
The green lines are contour lines indicating the
predicted values \GW{of the} cross section times
branching ratios for the channel in which
the excess was observed, i.e.~$\sigma(gg \to A)
\times \mathrm{BR}(A \to ZH)
\times \mathrm{BR}(H \to t \bar t)$.
In the green shaded areas the predicted
values are within the interval
given in \refeq{eq:excess}, corresponding to a description
of the excess within 
\GW{$1\,\sigma$}.

In both plots
one can see that for values of $1.2 \lesssim
\tan\beta \lesssim 1.8$
and \GW{values of $\cos(\beta - \alpha)$ near} the alignment limit
a description of the excess is possible
\GW{in accordance with the limits}
from searches
for additional scalars \jmn{and} 
\GW{with the measurements of the properties of the detected 
Higgs boson at 125~GeV}.
Other decay channels for~$H$, such as
$H \to VV$ and $H \to hh$, 
\GW{become relevant}
outside of the alignment limit.
As a result,
for a fixed value of $\tan\beta$
the predicted cross sections decrease
with increasing values of $|\cos(\beta - \alpha)|$.
This gives rise to the
the shape of the green lines peaking
at $\cos(\beta - \alpha) \approx 0$.
Regarding a \GW{possible}
description of the excess in combination with
deviations from the alignment limit, one can observe
in the left plot
that a sufficiently large cross section
of more than about 0.30~pb can be achieved for
$-0.15 \lesssim \cos(\beta - \alpha) \lesssim  \TBa{0.10}$, 
corresponding
to modifications of the couplings of~$h$ compared
to the SM predictions 
\GW{that can exceed}
10\%.
\TBa{However, for both types}
values of $|\cos(\beta - \alpha)|
\gtrsim \TBa{0.025}$ are 
\GW{excluded} by the cross
section measurements of the SM-like Higgs boson \TBa{ and/or by the
searches for $H \to hh \to b \bar b \tau^+ \tau^-$.
In type~II the exclusion regions
of both these constraints are largely
overlapping, as 
shown in the right
plot of \reffi{fig:excess}, where the condition
$\Delta \chi^2_{125} < 6.18$ yields an exclusion for values of
$\cos(\beta-\alpha) \lesssim -0.02$ and
$\cos(\beta-\alpha) \gtrsim 0.045$, while
the searches for $H \to hh$ exclude
$|\cos(\beta-\alpha)| \gtrsim 0.025$.
The situation is different in type~I,
where for negative values of
$\cos(\beta-\alpha)$ the parameter
region excluded
by \texttt{HiggsSignals} features values of
$\cos(\beta - \alpha) \lesssim -0.15$,
whereas the $H \to hh$ searches exclude
already values of $\cos(\beta - \alpha) \lesssim -0.025$
(as in type~II).}
\hto{Overall, we find that the observed excess can
readily
be accommodated by the 2HDM}
\htmm{within the level of $1\,\sigma$ while being in agreement with all cross-section limits
from BSM scalar searches and the 
measurements of 
\GW{properties of}
the detected Higgs boson
at~125~GeV.}

\subsubsection{Gravitational wave detection}

In the discussion in
\refse{sec:lowtanbeta} we 
already
\GW{pointed out}
that the considered benchmark point predicts
a FOEWPT according to the \htb{thermal
effective potential} approach.
\hto{The fact that the excess can be readily accommodated}
in the 2HDM, as
discussed above,
motivates a closer look at the FOEWPT
and related phenomenological consequences.
\hto{More concretely, we analyze}
whether the
stochastic GW background predicted by
a parameter point compatible with the excess could \GW{potentially} be
observed at future experiments,
in particular by LISA.

\begin{figure}[t]
\centering
\includegraphics[width=0.54\textwidth]{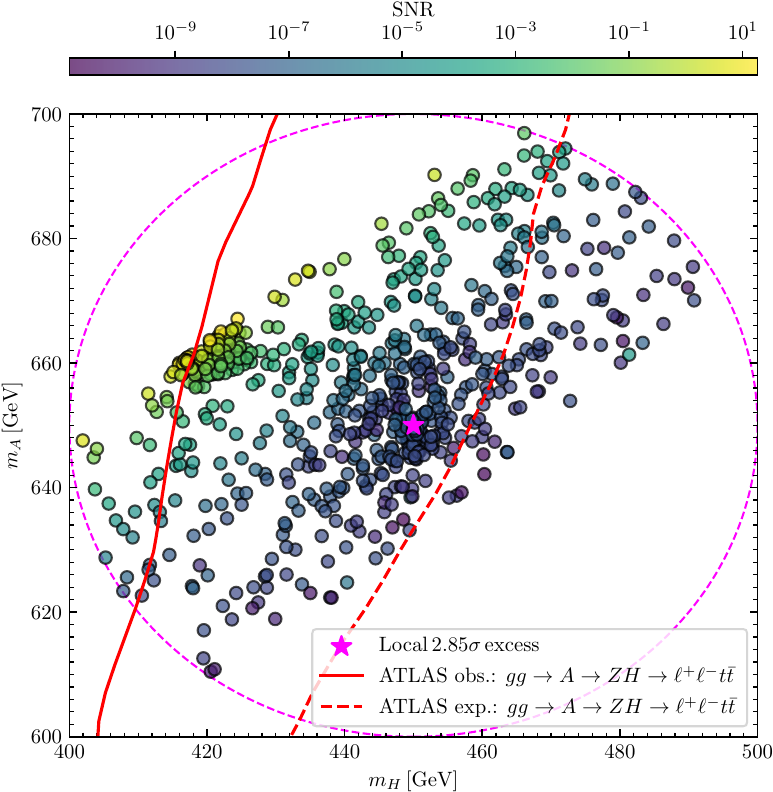}
\caption{\GW{The signal-to-noise ratio (SNR) of the scanned parameter 
points in the}
\htb{$(m_H , m_A)$-plane for
$\tan\beta=1.5$. \htmm{The mass values
for which the excess
observed by ATLAS in the $\ell^+ \ell^- t \bar t$
final state
was most pronounced}
($2.85\,\sigma$ local significance)
are indicated with a magenta star.
Shown are parameter points 
that feature a FOEWPT and a \GW{predicted} GW signal 
\GW{with a SNR at LISA that is} larger than $10^{-11}$.
\GW{We regard the region}
within the
\htmm{magenta dashed} 
\GW{contour as a rough estimate of the part of the parameter space 
that is}
compatible with
the description of the excess
\htmm{in view of} 
the mass resolution of the ATLAS search. \htmm{The expected and observed} exclusion
limits 
are \GW{indicated by the}
\htmm{red dashed and} solid lines,
\htmm{respectively.}
The colors \htmm{of the points 
\GW{display}
the SNRs of the GW signals
at LISA
assuming} 
an operation
time of seven years \htmm{and a bubble-wall
velocity of $v_{\rm w} = 0.6$}.}
}
\label{fig:snr}
\end{figure}

To this end, we \hto{performed a dedicated scan in $m_A$ and $m_H$
within a $\pm\,50$~GeV 
mass window around the} 
\GW{values of $m_A = 650\gev$ and $m_H = 450\gev$},
fixing the other 2HDM parameters as discussed
above\hto{. The \GW{prediction for the} GW signal was calculated for all}
the parameter points featuring a FOEWPT. 
In the 2HDM the
GW signals produced in FOEWPTs are very sensitive
to the \GW{precise values of the} scalar masses~\cite{Biekotter:2022kgf}.
Notably, a mere 50~GeV 
\GW{variation}
in the scalar masses
leads to a SNR spanning 
\GW{many}
orders of magnitude,
as can be seen in \reffi{fig:snr}, where we show
the parameter points \hto{featuring a FOEWPT
of the dedicated scan in the \htmm{($m_H$,$m_A$)-}plane}.
\GW{The}
colors of the points indicate the SNR of the GW signal
at LISA. For the computation of the SNR, we assume a
bubble wall velocity of $v_{\rm w} = 0.6$, and the LISA operation
time is assumed to be seven years,
see \refse{sec:gw} for details.
We only depict parameter points for which
SNR~$> 10^{-11}$.
One can see that the predicted values of the SNRs vary over
ten orders of magnitude within the \GW{relatively} 
small mass window
considered here.
If we consider as detectable GW signals
the ones with SNR~$\gtrsim 1$, the parameter space that 
can be probed with LISA is confined to 
the small region highlighted by the bright points.
\GW{The parameter region featuring the strongest FOEWPTs, which is 
indicated by the yellow points, is cut off by the onset of}
the phenomenon of
vacuum trapping, \GW{see the discussion in} \citere{Biekotter:2022kgf}. 
On the other hand, in the lower right area of the 
\GW{magenta dashed contour}
no points are present because
\GW{this parameter region gives rise to either a}
very weak FOEWPT \GW{yielding} GW signals
\GW{that are} far out of reach of LISA or
does not feature a first-order phase transition at all.

As a consequence of the \GW{limited} mass resolution
of the \GW{observed} excess and the very \GW{sensitive}
dependence
of the GW signals on the scalar masses,
no definitive conclusion can be
drawn about whether the 2HDM interpretation of the
excess would be associated with a primordial GW
\hto{signal} detectable with LISA.
We emphasize that this statement holds irrespective
of the substantial theoretical uncertainties present
in the computation of the phase transition parameters
and the GW signals (see
\citeres{Athron:2023xlk, Croon:2020cgk,Athron:2023rfq} for detailed 
\GW{discussions} of the theoretical uncertainties).
Even if 
\GW{the theoretical uncertainties from unknown higher-order contributions} 
\hto{in the calculation of the SNR}
\GW{were negligible},
the parametric uncertainties \GW{arising from the 
experimental error 
in the determination of the masses of the BSM Higgs bosons
would be a limiting factor for assessing the question whether}
\hto{such an excess}
would lead to
GW signals detectable with LISA or other future
GW detectors. 
\GW{We regard this finding as a generic feature}
of the interplay
between the LHC and GW detectors: while 
\GW{in the case of the} 
\htb{absence}
of \GW{a signal of BSM} physics at the LHC 
\GW{the resulting limits will place}
strong and definitive constraints
on the possibility of a detection of a GW signal produced
in a FOEWPT \GW{within the considered class of models}, 
a possible observation of BSM scalars 
\GW{may not provide sufficient information}
to make a clear prediction 
\GW{on} whether a GW detection
at LISA can be expected. 


\begin{figure}[t]
\centering
\includegraphics[width=0.54\textwidth]{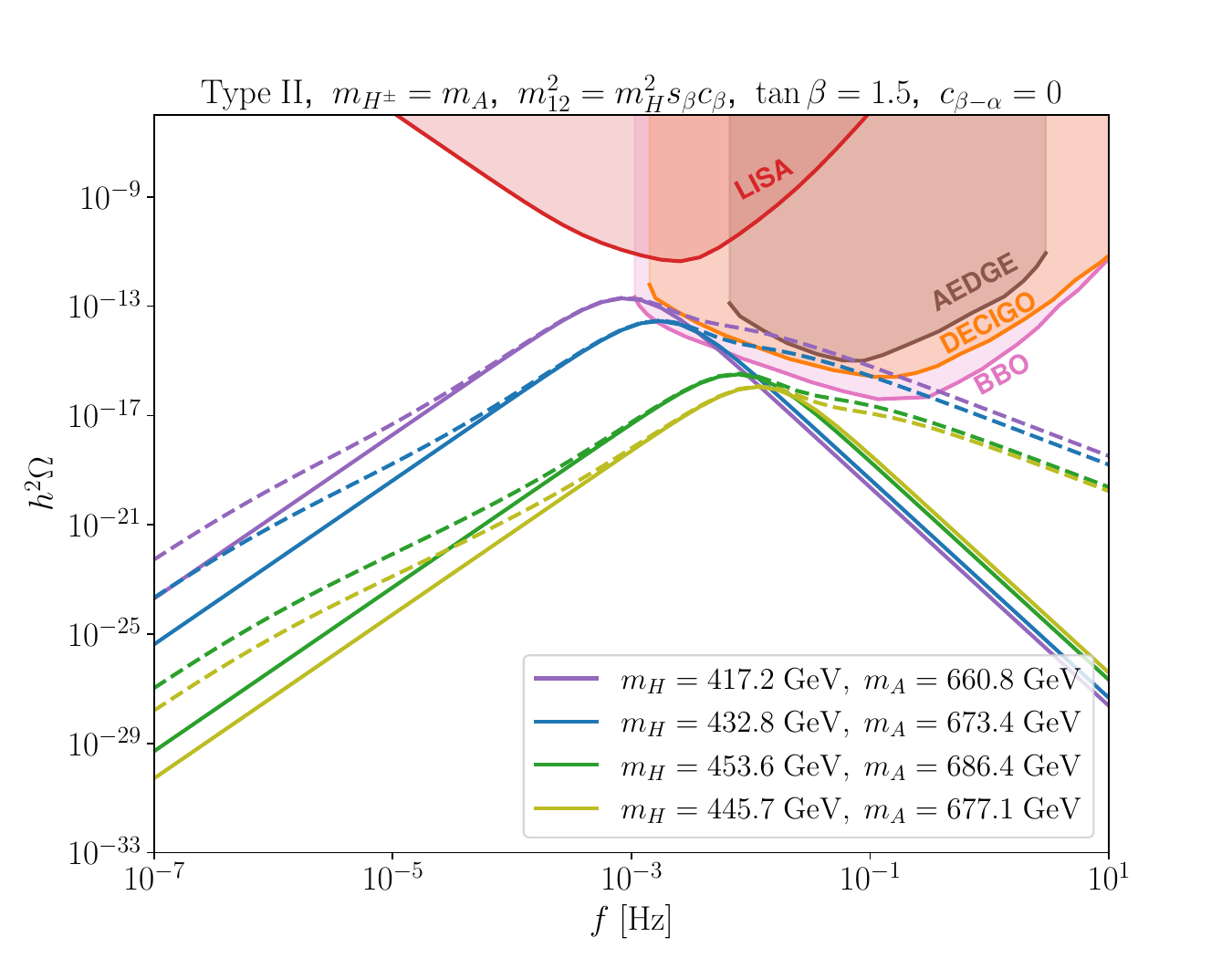}
\caption{Gravitational wave spectra for parameter
points \GW{specified in \refta{tab:ptparams} that are} compatible with the excess \GW{observed in the ATLAS search}. 
\GW{The solid} (dashed)
lines show \GW{the prediction without (including)}
the turbulence contribution, \GW{using $v_{\rm w} = 0.6$}. 
The colored regions show the \GW{prospective sensitivities} of future experiments.}
\label{fig:gwsignal}
\end{figure}

\begin{table}[t]
\centering
    \begin{tabular}{lll|rrrrr}
    	$m_H$ &  $m_A$& $m_A - m_H$ & $T_n$ & $v_n/T_n$  & $\alpha$ & $\beta/H$ & SNR \\
    	\hline
    	417.2 & 660.8 & 243.6 & 79.44  & 3.10  & 0.0308 & 77  & 13.7 \\
    	432.8 & 673.4 & 240.6 & 86.23  & 2.85  & 0.0206 & 134 & 3.8\\
    	453.6 & 686.4 & 232.8 & 110.89 & 2.19  & 0.0073 & 468 & 0.022\\
    	445.7 & 677.1 & 231.4 & 116.48 & 2.06  & 0.0062 & 674 & 0.004\\
        450.0 & 650.0 & 200.0 & 145.08 & 1.42  & 0.0029 & 5399 & $< 0.001$\\
    \end{tabular}
    \caption{\GW{Results for parameters characterizing}
    the phase transition for 
    \GW{example points of the 2HDM that are} compatible with the excess 
    \GW{observed in the ATLAS search. The corresponding}
    GW spectra \GW{are} shown in Fig.~\ref{fig:gwsignal}.
    \htmm{Dimensionful parameters are
    given in~GeV.}
    \htg{The SNR values \hto{evaluated for LISA} include the turbulence contribution.}
    }
       \label{tab:ptparams}
\end{table}

To further illustrate the impact of the
experimental mass resolution of BSM scalar
searches at the LHC on the predicted GW
signals, we show in \reffi{fig:gwsignal}
the spectral shape of the GW backgrounds
produced during 
\GW{a} FOEWPT for \htg{several parameter 
points with masses of the heavy scalars
specified in Tab.~\ref{tab:ptparams}
together
with the parameters that characterize
the phase transition. The}
remaining 2HDM parameters
are kept fixed
according to the previous discussion. \htg{ We chose the point with the largest SNR found in \reffi{fig:snr} and allow for up to 10\% deviations in the values of the masses $m_H$, $m_A$, which translates into deviations of the SNR of several orders of magnitude.
\hto{In addition, we show} \GW{in \refta{tab:ptparams}} the parameters for the point $(m_H,m_A) = (450,650)$ GeV although we omit its GW spectrum 
\GW{in \reffi{fig:gwsignal}}
because of the smallness of the SNR.}
The spectral shapes of the GW backgrounds
are computed as discussed in \refse{sec:gw},
where the solid curves depict the sound-wave
contribution $h^2 \Omega_{\rm sw}$
only, whereas the dashed curves
depict the sum of sound-wave 
\GW{and}
turbulence 
contributions, i.e.~$h^2 \Omega_{\rm sw} +
h^2 \Omega_{\rm turb}$.
We also show the sensitivity curves
of LISA~\cite{LISACosmologyWorkingGroup:2022jok}, AEDGE~\cite{AEDGE:2019nxb}, DECIGO~\cite{Kawamura:2011zz,Kawamura:2020pcg} and BBO~\cite{Corbin:2005ny}, \hto{where the latter three} are planned\hto{,
but not yet approved}
space-based GW detectors.
One can see that only for the smallest value \GW{of}
\GW{$m_H = 417.2\gev$},
i.e.~the largest mass splitting
between $H$ and $A$, the GW signal might
be detectable with LISA, according to the predicted SNR.
For values of $m_H$ only
a few percent larger, the peak amplitudes of the
GW signals \GW{drastically decrease and quickly
drop to values} far below the experimental sensitivity
of the proposed GW detectors.
We emphasize \GW{again at this point} that the detectability of the
GW signal for a single parameter point cannot
be determined definitively with the methods
applied here due to the 
\GW{substantial}
theoretical
uncertainties in the prediction of the GW
signals. However, the fact that 
\hto{in the case of a possible detection of BSM scalars at the LHC}
a 
mass resolution \hto{at the percent level}
would be required 
in order to draw \hto{conclusions about the detectability of
a GW signal} 
\GW{poses a challenge independently of the status of the remaining}
theoretical uncertainties \GW{at that time}.

Of course, one can also turn this argument
around. An LHC discovery, e.g.~a signal in the smoking-gun signature, in combination with a GW detection
at LISA that is consistent with a FOEWPT 
\hto{as interpreted in a}
UV-complete model, could be used for a more
precise (but model-dependent)
determination \hto{of the parameters of 
\GW{the considered}
BSM Higgs sector.}
In this way space-based GW astronomy \jmn{could}
become a complementary tool to sharpen the precision of particle physics.\footnote{\jmn{This would be similar in spirit 
to the present situation regarding the 
sum of neutrino masses, constrained
most stringently using astrophysical observations,
e.g.~the measurement of the spectrum
of the cosmic microwave
background~\cite{DiValentino:2021hoh}.}
}

\section{Summary and conclusions}
\label{sec:conclus}

Recently, ATLAS has reported for the first time, 
\GW{and} \hto{based on} the full Run~2 data set collected
at~13~TeV, 
the results for searches for additional Higgs bosons
\GW{where a}
heavier \hto{(pseudo-)}scalar resonance~$A$
\GW{is produced} via gluon-fusion and subsequently
decays into a $Z$ boson
and a lighter scalar resonance~$H$\hto{. The} 
\GW{search made use of the leptonic decay of the}
$Z$ boson,
\GW{while for} the lighter
scalar the \GW{search focused on the} decay into 
\hto{a} top-quark pair.
This signature is exceptionally promising \GW{for probing}
the 2HDM parameter space 
\GW{for the case where the masses of the neutral BSM scalars are}
above the di-top threshold and 
\GW{have a splitting that is at least as large as the mass of the $Z$ 
boson}.
Consequently, \hto{within the 2HDM}
this signature has been identified as a smoking-gun
signature for a FOEWPT, whose presence relies on sizable mass splittings
in order to generate a potential barrier separating
the symmetry-conserving and the symmetry-breaking
vacua. \GW{In this context in particular the region of} 
low $\tan\beta$ \GW{is of interest, which is}
preferred by EW baryogenesis.
Since the searches in the $\ell^+ \ell^- t \bar t$
final state are able
to probe 
parameter space regions of the~2HDM
\GW{that were unconstrained up to now},
we \GW{have performed} \hto{a comprehensive} 
analysis of the impact of the cross-section
limits reported by~ATLAS. We focused on 2HDM benchmark scenarios assuming the alignment limit,
$\cos(\beta - \alpha) = 0$, mass degeneracy between
the charged Higgs boson and the pseudoscalar state,
$m_{H^\pm} = m_A$, and setting the decoupling
scale $M$ (defined by the relation
$m_{12}^2 = M^2 \sin\beta \cos\beta$)
equal to the mass of the 
\GW{heavier}
CP-even scalar,
$M = m_H$.

In the first part of our analysis, we determined
the parameter regions \GW{that are} excluded by this
new search at the 95\%~C.L.~in
the $(m_H,m_A)$ plane.
We started by considering a low-$\tan\beta$ regime
with $1 \leq \tan\beta \leq 3$,
\GW{preferred in view of a possible realization of EW 
baryogenesis and associated with relatively large couplings of $A$ and $H$ to top quarks, implying that}
the new ATLAS $\ell^+ \ell^- t \bar t$
\GW{has high sensitivity}.
We found that for $\tan\beta = 1$ the new search
exclude\hto{s} a large region of parameter space that so
far was not constrained by LHC searches.
In combination with LHC limits from
searches for $H^\pm \to tb$, $H \to \tau^+ \tau^-$
and $H,A \to t \bar t$, masses of $300\gev \lesssim
m_H \lesssim 450\gev$ are now entirely excluded
\GW{in this scenario},
while previously \hto{for $\tan\beta = 1$} a wide parameter space region
with $350\gev \lesssim m_H \lesssim 460\gev$
and $650\gev \lesssim m_A \lesssim 800\gev$ was
left unconstrained. 
\GW{For increasing values} of~$\tan\beta$,
the charged scalar searches lose sensitivity, \jmn{and in particular}
for $\tan\beta = 1.5$
the $\ell^+ \ell^- t \bar t$ search is the only
LHC search that can currently
probe the parameter space regions
\jmn{featuring} 
a~FOEWPT
above the di-top threshold, $m_H > 2\, m_t$.
\GW{For} $\tan\beta = 2$,
and irrespectively of whether the presence
of a~FOEWPT is required or not, the
search for the smoking gun signature $A \to ZH$
with $H \to t \bar t$ is currently the
only LHC search \GW{that is} able
to exclude parameter space regions in our
benchmark plane with~$m_H$ above
the di-top threshold. For $\tan\beta = 3$, the largest
value considered in the low-$\tan\beta$ regime,
the $\ell^+ \ell^- t \bar t$ searches are currently
not 
\GW{yet} able to
probe the 2HDM parameter plane, because
the gluon-fusion production cross section
of~$A$ is too small. Instead, we demonstrated that
searches for $A \to ZH$ with $H \to b \bar b$,
which had previously been carried out by both
ATLAS and CMS including the full Run~2 datasets, are
more promising \GW{in this scenario}, giving rise to an exclusion region
reaching masses of up to $m_H \approx 400\gev$.
\GW{We have furthermore pointed out}
a novel, not yet performed LHC search that would be complementary to the smoking gun $A \to ZH$ search 
\GW{in the low $\tan\beta$ region}
as a probe of the 2HDM parameter region featuring a strong FOEWPT.
\GW{The channel that we propose for experimental analysis consists of}
$H^{\pm}$ production (e.g.\ via $p p \to H^{\pm} t b$) followed by the 
decay $H^\pm \to W^\pm H \to \ell^{\pm} \nu \,t \bar t$.

In addition to
the searches in the $\ell^+ \ell^- t \bar t$
final state, ATLAS also reported for the first
time searches for the $A \to ZH$ decay 
\GW{making use of}
the decay of the $Z$ boson into neutrino pairs \jmn{and the  $H \to b \bar b$ decay mode}.
Here, both the gluon-fusion and the $b \bar b$-associated
production of~$A$ were considered.
These searches in the $\nu \nu b \bar b$ final state
\jmn{may become important for large} values of~$\tan\beta$.
\GW{Investigating} a representative
2HDM benchmark scenario with $\tan\beta = 10$ (and the
remaining 2HDM parameters as described above), we found
that in the 2HDM the earlier searches in the
$\ell^+ \ell^- b \bar b$ final state give rise to
stronger exclusions than the new \jmn{$\nu \nu b \bar b$ ATLAS search.}
Nevertheless, it should be taken into account that
the new searches utilizing the $Z \to \nu \nu$
decay mode cover a wider mass interval
and larger mass splittings between the two BSM
resonances.
In the 2HDM the maximum amount of
mass splittings between the BSM
scalars is limited by the perturbativity constraint: the scalars are contained in the same
SU(2) doublet, such that their masses are
confined to
lie not too far away from the overall decoupling
scale~$M$. In other models, in which additional
mass scales are present and larger mass splittings
can be realized between different BSM scalars,
the new searches in the
$\nu \nu b \bar b$ final state could be able to
probe so far unconstrained parameter space regions.


\vspace{1mm}

\GW{As a further part of our analysis,}
motivated by the strong impact of the 
$\ell^+ \ell^- t \bar t$ searches \jmn{described above},
we investigated in \refse{sec:prospects} the future prospects in terms of projected exclusion regions, which were obtained via a simple rescaling of the reported
expected cross-section limits of the new ATLAS search with
integrated luminosities anticipated to be collected in the future at the (HL-)LHC.
We found that the 
\GW{reach} of the smoking gun signature will significantly improve
and parameter regions with \jmn{$m_H > 2\, m_t$ and $\tan\beta \geq 3$} will become accessible. 
We therefore anticipate that the smoking gun signature $A \to Z H$ with $H \to t \bar t$ decay will be the main LHC search channel in the future to probe the 
\GW{allowed} 2HDM
parameter space regions above the di-top threshold that \jmn{feature} a strong FOEWPT.
\jmn{Besides,} given that successful EW baryogenesis prefers small $\tan\beta$ values (not
much larger than one)~\cite{Fromme:2006wx}, \jmn{such searches} 
will have a large impact on the possibility of explaining the matter-antimatter asymmetry
\jmn{via} EW baryogenesis \jmn{in the 2HDM}. 
%

\vspace{1mm}

\jmn{The new ATLAS search in the $\ell^+ \ell^- t \bar t$ final state showed an excess which} 
\jmn{is most significant for masses of $m_A = 650\gev$ and $m_H = 450\gev$, with a local significance of $2.85\,\sigma$. We have demonstrated that the excess can be described at the~$1\,\sigma$ level
in the 2HDM \hto{type~I and~II (as representative scenarios)} in the approximate range 
\GW{of \TBn{$\tan\beta$ values} between 1.2 and 1.8}
for the alignment limit (a $1\,\sigma$ description of the excess is also possible for small departures from the alignment limit, with the viable $\tan\beta$ range shrinking accordingly), while being in agreement with all existing bounds from BSM scalar searches at the LHC. 
Further probes of the nature of the excess could be performed by searching for the decay 
$H^\pm \to W^\pm H$ followed by $H \to t \bar t$, as discussed above.}
Notably, the masses corresponding to the excess lie within
the parameter space region indicative of a strong~FOEWPT based on the
thermal effective potential approach applied here. 
This parameter region could thus be suitable for successful EW baryogenesis. 
We emphasize, however, that in order to investigate whether
the BAU can be predicted in agreement with observations one would
have to take into account 
\GW{additional} sources of CP-violation, required for the generation
of the~BAU according to the Sakharov conditions. These sources of CP-violation might have an
impact on the cross section for the process
in which the excess was observed.
We leave a \GW{more detailed} investigation of the predicted
BAU and \GW{of the possible impact of new sources of} CP-violation on the
description of the excess for future studies.
We analyzed the primordial GW \jmn{signal that}
would be generated during the FOEWPT in the 2HDM parameter space region compatible
with the observed excess. \jmn{We found that, since the} 
predictions for the GW spectra are highly sensitive to the precise values of
the BSM scalar masses, 
\GW{the signal-to-noise ratio expected at LISA varies by}
\jmn{
several order of magnitude 
for points within the region compatible with the ATLAS excess}.\footnote{We stress again that this is purely 
\GW{a consequence of}
the parametric uncertainty stemming from
the experimental mass resolution of BSM scalar searches at the LHC, 
\GW{which will pose a challenge independently of whether the}
theoretical
uncertainties on the predictions for the GW
power spectra 
\GW{arising from higher-order effects can significantly be reduced}.}
\GW{Therefore at this stage}
no definitive statement can be made about whether the GW backgrounds
would be detectable at LISA (or other future space-based GW detectors).
\jmn{Nevertheless, should a stochastic GW signal be detected by LISA,} 
\GW{in combination with an LHC signal as hinted by the ATLAS excess 
this would within the context of the 2HDM provide new and very precise information on the allowed values of the parameters.}


\GW{In} order to produce the results presented in this paper, we developed the software package \texttt{thdmTools}, which can be used for the phenomenological investigation
of the CP-conserving 2HDM with softly-broken  $\mathbb{Z}_2$ symmetry.
Accompanying this paper, we make the code \texttt{thdmTools} available to the
public. Installation instructions and a brief discussion about the functionalities of the
code and its interfaces to other public codes, \htmm{in particular to \texttt{HiggsTools}} 
are given in \refap{app:code}.

\medskip

\jmn{
\GW{Our} analysis shows that the smoking gun signature $A \to ZH$
(with $H \to t \bar t$) has great potential to further probe the viable 2HDM parameter regions, in particular those that may feature a strong FOEWPT as required for EW baryogenesis. 
\GW{In fact, the excess that has been observed in this search could be}
the first hint of such a strong transition in the 2HDM.}


\section*{Acknowledgements}

We thank Philipp Gadow, Daniel Hundhausen, Benoit Laurent, Matthias Schröder and Matthias Steinhauser for helpful discussions.
The work of T.B.~is supported by the German
Bundesministerium f\"ur Bildung und Forschung (BMBF, Federal
Ministry of Education and Research) – project 05H21VKCCA.
The work of M.O.O.R~is supported by the European Union's
Horizon 2020 research and innovation programme under grant agreement No 101002846, ERC CoG ``CosmoChart''.
J.M.N.\ was supported by the Ram\'on y Cajal Fellowship contract RYC-2017-22986, and acknowledges partial financial support by  the grant PID2021-124704NB-I00, and by the European Union's Horizon 2020 research and innovation programme under the Marie Sk\l odowska-Curie grant agreements No 860881-HIDDeN and 101086085-ASYMMETRY. 
\hto{S.H.\ and J.M.N.\ acknowledge partial financial support by the Spanish Research Agency (Agencia Estatal de Investigaci\'on) through the grant IFT Centro de Excelencia Severo Ochoa No CEX2020-001007-S funded by 
MCIN/AEI/10.13039/501100011033. The work of S.H.\ was supported in part by the
grant PID2019-110058GB-C21 funded by MCIN/AEI/10.13039/501100011033 and by
``ERDF A way of making Europe'', and in part by the Grant PID2022-142545NB-C21 funded by
MCIN/AEI/10.13039/501100011033/ FEDER, UE.}
K.R. and G.W acknowledge support by the
Deutsche Forschungsgemeinschaft (DFG, German Research Foundation) under Germany’s
Excellence Strategy~–~EXC 2121 “Quantum Universe”~–~390833306. This work has been partially funded by the Deutsche Forschungsgemeinschaft (DFG, German Research Foundation)~-~491245950.

\appendix

\section{The python package \texttt{thdmTools}}
\label{app:code}

\texttt{thdmTools} is a python package
\GW{\TBn{for} exploring} the 
Two Higgs Doublet Model
\GW{with real parameters and a}
softly-broken $\mathbb{Z}_2$ symmetry,
\GW{which incorporates tests of}
the relevant theoretical
and experimental constraints. 
It allows the user to specify a parameter
point in terms of the free parameters of
the model (see \refeq{eq:freeparas}).
During the installation of this package
the following external codes will also be downloaded and installed:
\begin{description}[font=\normalfont]
\item[\texttt{AnyHdecay}\cite{Muhlleitner:2016mzt,Djouadi:1997yw,Djouadi:2018xqq}:] Computes the branching ratios and decay widths of all
Higgs bosons contained in the model
\item[\texttt{HiggsTools}\cite{Bahl:2022igd}:]
    Checks compatibility with the experimental
    constraints on BSM scalars from LEP and
    LHC searches 
    (\texttt{HiggsBounds}~\cite{Bechtle:2008jh,
Bechtle:2011sb,Bechtle:2013wla,
Bechtle:2020pkv})
    and with existing measurements of the
    detected \GW{Higgs boson at about} 125~GeV
    (\texttt{HiggsSignals}~\cite{Bechtle:2013xfa,
Bechtle:2014ewa,Bechtle:2020uwn}).
\item[\texttt{THDM\_EWPOS}
    \cite{Hessenberger:2016atw,
    HessenbergerPhD,
    Hessenberger:2022tcx}:] Computes the prediction for some Electroweak Precision Observables (EWPO), in particular $M_W$ \GW{and the effective weak mixing angle at the $Z$-boson resonance}, in the
    2HDM at the two-loop level.
\item[\texttt{SuperIso}
    \cite{Mahmoudi:2007vz,Mahmoudi:2008tp}:]
    Computes the predictions for
    various flavour-physics observables
    in the 2HDM.
\end{description}
Note that \texttt{HiggsTools} will only be
installed if not already installed in the
current \texttt{python} environment of the
user.

\subsection{Installation}
\texttt{thdmTools} is publicly available at:
\begin{center}
\url{https://gitlab.com/thdmtools}
\end{center}
Installation requires a \texttt{python3} environment and compilers for \texttt{Fortran} and \texttt{C++}.
The package and all its dependencies
are installed by executing:

\begin{minted}[bgcolor=bg]{python}
make all
\end{minted}

\subsection{Code example for thdmTools}

To import the package in an example notebook type
\begin{minted}[bgcolor=bg]{python}
from thdmTools.parampoint import Parampoint
\end{minted}

\noindent \GW{To} define the parameters of the point as a dictionary in python 
\begin{minted}[bgcolor=bg]{python}
dc = {
    'type': 1,
    'tb': 3,
    'alpha': -0.3217,
    'mHh': 500,
    'mHl': 125.09,
    'mA': 500,
    'mHp': 500,
    'm12sq': 75000}
\end{minted}

\noindent \GW{To} run the code and check all the included theoretical and experimental constraints

\begin{minted}[bgcolor=bg]{python}
pt = Parampoint(dc)
print('Stability:', pt.check_vacuum_stability())
print('Unitarity:', pt.check_perturbative_unitarity())
print('Flavour:',pt.check_flavour_constraints())
print('EWPO:', pt.check_ewpo_constraints())
print('Collider (HB,HS):', pt.check_collider_constraints())
\end{minted}
\GW{These} commands will print a boolean statement of True/False if the point is allowed/disallowed by the corresponding constraint\htg{, see below for further details on the criteria applied in each case.}
Vacuum stability can be checked with the conditions
on boundedness from below of the tree level potential,
see \GW{e.g.}~\citere{Bhattacharyya:2015nca}.
An additional condition that
ensures that the EW minimum
is the global minimum of the tree level
potential~\cite{Barroso:2013awa} can be applied
by setting the optional argument
\texttt{global_min} of the
function \texttt{check_vacuum_stability()}
to \texttt{True}:
\begin{minted}[bgcolor=bg]{python}
print('Stability:', pt.check_vacuum_stability(global_min=True))
\end{minted}
In order to 
\GW{test}
perturbative unitarity the default configuration 
\GW{checks} for the unitarity of the scalar four-point scattering matrix in the high-energy limit, applying an
upper limit of~$8\pi$ to the eigenvalues
of the scattering matrix.
One can change the upper limit by
giving an argument in the function \begin{minted}[bgcolor=bg]{python}
('Unitarity to 1:', pt.check_perturbative_unitarity(1.))
\end{minted}
where the value of the argument multiplied
by~$16\pi$ is the applied upper limit
(thus 0.5 is the default value).
One can also check for NLO perturbative unitarity in the high enery limit according to the expressions derived in \citere{Cacchio:2016qyh}:
\begin{minted}[bgcolor=bg]{python}
('NLO Unitarity:', pt.check_perturbative_unitarity_nlo())
('NLO Unitarity to 1:', pt.check_perturbative_unitarity_nlo(1.))
\end{minted}

\noindent 
\GW{Regarding constraints from
flavor physics the predictions} for $B \rightarrow s\gamma$ and $B_s \rightarrow \mu^+\mu^-$ are checked. One can access the 
\GW{specified allowed region}
by typing
\begin{minted}[bgcolor=bg]{python}
print(pt.b2sgam_valid)
print(pt.bs2mumu_valid)
\end{minted}
Many other predictions for 
\GW{flavor-physics}
observables are available via the interface
to \texttt{SuperIso}, such that 
\GW{users
can define their} 
own functions to exclude
or accept parameter points including additional
observables.

The function called above to check against
constraints from EWPO
calls the interface to \texttt{THDM\_EWPOS}.
\GW{This function in particular} verifies whether
the predicted values for the $W$-boson mass,
\htmm{the total decay width of the $Z$ boson
and the effective weak mixing angle} \GW{at the $Z$-boson resonance}
are in agreement with the 
\htmm{experimental measurements}
(\htmm{by default the prediction for
$M_W$ is checked against the average
value from the 
\GW{LHC--TeV $M_W$ Working Group}~\cite{Amoroso:2023pey},
\GW{which does not include} the CDF $M_W$ result})
within two standard deviations.
The experimental values and their uncertainties
can be changed by the user via optional
arguments to the function
\texttt{check_ewpo_constraints()}.
%
As 
\GW{a further possibility,}
the user can perform a $\chi^2$ fit to the experimental
EWPO in terms of the oblique
parameters $S$, $T$, and~$U$ by typing
\begin{minted}[bgcolor=bg]{python}
print('STU chi^2 fit:' pt.check_ewpo_fit())
\end{minted}
\GW{In contrast to the analysis of the EWPO as specified above, 
the $S$, $T$, $U$ parameters are evaluated only at the}
one-loop level according to
\hto{\citere{Grimus:2007if}}.  \GW{The} experimental
fit values of the oblique parameters,
their uncertainties and correlations are
taken from \citere{Haller:2018nnx}.
By default, a parameter point is considered
to be viable \GW{according to this approach} 
if the predicted values of
the $S$, $T$, $U$ parameters are in agreement
with the experimental fit values within the
$2\,\sigma$ confidence level.

\GW{Regarding} the collider constraints, for compatibility with the 125 GeV Higgs-boson measurements one can specify the 
\GW{allowed range of}
\htg{$\Delta\chi^2_{125}$} to 
\GW{perform a test using}
the best $\chi^2$ fit performed with \texttt{HiggsSignals}. The default is to allow
parameter points with \htg{$\chi^2_{125}$}-values that are not more than $2\,\sigma$ away from the SM \htg{$\chi^2_{125}$}-value, which 
\GW{corresponds to}
\htg{$\Delta\chi^2_{125}= 6.18$}
in two-dimensional parameter 
\GW{representations}.
For compatibility with cross-section limits
from searches for additional scalars,
one can 
call the \texttt{HiggsBounds} interface
and access the result 
by means of
\begin{minted}[bgcolor=bg]{python}
print('HiggsBounds result:', pt.reshb)
\end{minted}
which will print a table with the observed/expected ratios of the most sensitive channels for each of the scalars in the~2HDM. 
The full information of the \texttt{HiggsBounds}
analysis is stored in the object
\texttt{pt.reshb} and can be accessed using
the various functionalities of
\texttt{HiggsTools}~\cite{Bahl:2022igd}.
The \GW{cross-section  
predictions from}
\texttt{HiggsTools}
for the LHC operating at~13~TeV
(based on the effective coupling input)
can also be accessed by typing, e.g.
\begin{minted}[bgcolor=bg]{python}
print('sigma(gg -> h)', pt.XS['Hl']['gg'])
\end{minted}
where \texttt{Hl} refers to the lighter CP-even
scalar and \texttt{gg} selects the gluon-fusion
production cross section.
The cross sections of the
second CP-even scalar, the CP-odd scalar and
the charged scalars can be chosen by using the keys
\texttt{Hh}, \texttt{A} and \texttt{Hp}, respectively.
For the neutral states, the other available production
modes stored in \texttt{pt.XS} are
$b \bar b$-associated production, vector-boson fusion,
$t \bar t$-associated production and
$t$-associated production, which can be accessed
with the keys \texttt{bbH}, \texttt{vbfH}, \texttt{ttH}
and \texttt{tH}, respectively.
Finally, it is also possible to access all branching ratios and total decay widths of the particles by typing
\begin{minted}[bgcolor=bg]{python}
pt.calculate_branching_ratios()
print('BR_h', pt.b_Hl)
print('BR_H', pt.b_Hh)
print('BR_A', pt.b_A)
print('BR_Hp', pt.b_Hp)
print('Total decay width h:', pt.wTot['Hl'])
print('Total decay width H:', pt.wTot['Hh'])
print('Total decay width A:', pt.wTot['A'])
print('Total decay width Hp:', pt.wTot['Hp'])
\end{minted}
where the predictions for the branching ratios
are computed via the interface to the
\texttt{AnyHdecay} library.
\label{app:tools}

\section{Comparison to previous CMS projections}
\label{app:proj}

\begin{figure}[h!]
\centering
\includegraphics[width=0.48\textwidth]{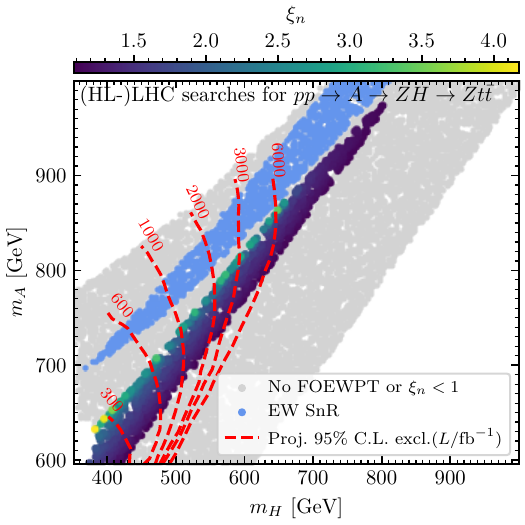}~
\includegraphics[width=0.48\textwidth]{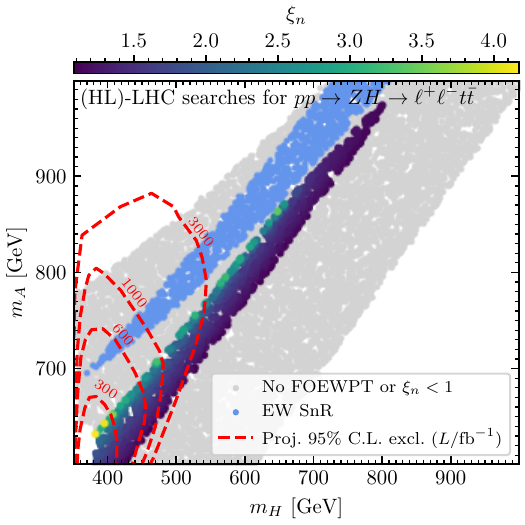}
\caption{Projected
exclusion regions 
\TBa{in the $(m_H,m_A)$ plane with $\tan\beta = 3$
and $m_{H^\pm} = m_A$ and}
\GW{for} integrated luminosities
of $300,600,1000,3000~\mathrm{fb}^{-1}$, expected
to be collected in future runs of the LHC. \GW{The displayed limits are} derived from rescaled CMS (left) and ATLAS (right) expected limits for the $\ell^+ \ell^- t \bar t$ final state. The color bar indicates the strength of the phase transition. The blue points indicate the parameter region that features electroweak symmetry non-restoration at high temperatures (see Ref.~\cite{Biekotter:2022kgf} for more details).}
\label{fig:cmsverifi}
\end{figure}

\htb{In Ref.~\cite{Biekotter:2022kgf}, we estimated the projected (HL-)LHC sensitivity 
\GW{for}
the process $A \to ZH$ in the $Z\, t \bar t$ final state 
for several integrated luminosities. We used the 
results 
\GW{for the expected sensitivity in}
this channel
obtained in a Master thesis 
for the CMS Collaboration~\cite{dpgcms,fischerthesis}
\TBa{and applied them to the $(m_H,m_A)$ parameter plane
also investigated in this paper, with $\tan\beta = 3$
and $m_{H^\pm} = m_A$}.
Here we aim to compare these prior sensitivity projections with those 
\GW{based on the} ATLAS expected limits, as detailed in Section~\ref{sec:prospects}.
In Fig.~\ref{fig:cmsverifi} we present the
\htmm{resulting} expected $95\%$ C.L.~exclusion \htmm{limits} 
corresponding to integrated luminosities of $300, \, 600, \, 1000, \, 3000~\mathrm{fb}^{-1}$, projected for future (HL-) LHC runs. On the left-hand side, we display the exclusion regions derived from a straightforward rescaling of the CMS expected limits for different luminosity values,
\htmm{thus not accounting for changes
in systematic uncertainties}.
On the right-hand side we show the exclusion regions resulting from a similar rescaling process, albeit based on ATLAS expected limits\htmm{, again without accounting
for possible changes in systematic uncertainties.}
The color code shows the phase transition strength $\xi_n$
for parameter points featuring
a FOEWPT with $\xi_n > 1$.
The blue region indicates the area that features electroweak symmetry non-restoration at high temperatures (see \citere{Biekotter:2022kgf} for details).
The comparison \hto{demonstrates} 
\htmm{good} agreement between both sets of projections, reinforcing the robustness of our conservative estimate of the future prospects.}

\bibliographystyle{JHEP}
\bibliography{lit}

\end{document}